\documentclass[10pt,letterpaper]{article}
\usepackage[top=0.85in,left=1.00in,right=1.00in,footskip=0.75in,marginparwidth=2in]{geometry}

\usepackage[utf8]{inputenc}

\usepackage{cite}

\usepackage{nameref,hyperref}

\usepackage{microtype}
\DisableLigatures[f]{encoding = *, family = * }

\raggedright
\usepackage{changepage}

\usepackage[aboveskip=1pt,labelfont=bf,labelsep=period,singlelinecheck=off]{caption}

\makeatletter
\renewcommand{\@biblabel}[1]{\quad#1.}
\makeatother

\usepackage{lastpage,fancyhdr,graphicx}
\usepackage{epstopdf}
\fancyheadoffset[L]{2.25in}
\fancyfootoffset[L]{2.25in}

\usepackage{color}

\definecolor{Gray}{gray}{.25}

\usepackage{graphicx}

\usepackage{sidecap}

\usepackage{wrapfig}
\usepackage[pscoord]{eso-pic}
\usepackage[fulladjust]{marginnote}
\reversemarginpar
\usepackage{amsmath}
\usepackage{bm}
\usepackage[maxfloats=256]{morefloats}
\usepackage{graphicx}
\usepackage{dcolumn}
\usepackage{tabularx}
\usepackage{multirow}
\usepackage{ulem}
\newcolumntype{M}{>{\centering\arraybackslash}m{1.85cm}}
\usepackage[export]{adjustbox}
\usepackage{float}
\usepackage{color}   
\usepackage{arydshln}
\newcommand{\keywords}[1]{\textbf{Keywords:} #1}
\usepackage{appendix}
\usepackage{mathtools}
\newcommand{\acknowledgments}[1]{\par \textbf{Acknowledgments:} #1}
\allowdisplaybreaks
\begin{document}
\justify
\vspace*{0.35in}

\begin{flushleft}
{\Large
\textbf\newline{Isospin symmetry breaking in atomic nuclei}
}
\newline
\\
J. A. Sheikh\textsuperscript{1,2,*},
S. P. Rouoof\textsuperscript{2},
R. N. Ali\textsuperscript{3},
N. Rather\textsuperscript{2,*},
Chandan Sarma\textsuperscript{4},
P. C. Srivastava\textsuperscript{4},
\\
\bigskip
\bf{1} Department of Physics, University of Kashmir, Srinagar, 190 006, India
\\
\bf{2}  Department of  Physics, Islamic University of Science and Technology, Awantipora, 192 122, India
\\
\bf{3} Department of Physics, Central University of Kashmir, Tulmulla, Ganderbal, 191131, India
\\
\bf{4} Department of Physics, Indian Institute of Technology Roorkee, Roorkee, 247667, India
\\
\bigskip
* sjaphysics@gmail.com (JAS), qadriniyaz@gmail.com (NR)

\end{flushleft}

\section*{Abstract}
The importance of the isospin symmetry and its breaking in elucidating the properties of atomic
  nuclei is reviewed. The quark mass splitting and the electromagnetic origin of the isospin symmetry breaking (ISB) for
  nuclear many-body problem is discussed. The experimental data on isobaric analogue states cannot be described only with the Coulomb interaction, and ISB terms in the nucleon-nucleon interaction are needed to discern the observed properties. In the present work, the ISB terms are explicitly considered in nuclear energy density functional and spherical shell model approaches, and a detailed investigation of the analogue states and other properties of nuclei is performed. It is observed that isospin mixing is largest
  for the $N=Z$ system in the density functional approach.
  \\
\keywords{Shell Model, Density Functional Theory, Isospin Symmetry, Charge Symmetry Breaking, Isobaric Analog States, 
Isobaric Mass Multiplet Equation.}   



\tableofcontents
\section{Introduction} \label{sec:intro}

The word "symmetry" is a 16th century Latin derivative from the Greek words for "syn-" (together) and "metron" (measure).
Symmetries have played a pivotal role in the development of the physical laws and breaking of these symmetries
gives rise to many phenomena occurring in nature. In the development of three monumental theories of physics,
the Newtonian mechanics, Einstein's relativity and the Maxwell's electromagnetism, symmetries played a prominent role. The concepts
of Galilean and Lorentz invariance principles, equivalence of inertial and gravitational masses and the duality between electric
and magnetic fields are the building blocks of these theories. In quantum mechanics, the symmetries are of paramount importance
as they lead to conserved quantum numbers which allow the characterization of the quantum mechanical states. The fundamental
translational and rotational symmetries in space-time lead to energy, momentum and angular-momentum quantum numbers
\cite{Weinberg_1995}. These symmetries have classical manifestations, but there is another class of symmetries that have no
classical analogies and occur only in the quantum mechanical realm. These include exchange symmetry
and the symmetries that are broken spontaneously. The former leads to the Pauli exclusion principle for fermions, and the
latter is an important class of symmetries associated with phase transitions giving rise to the Nambu-Goldstone
boson \cite{Nambu1960, Goldstone1961}.

The main focus of the present work is to review, and also to discuss some new results on
isospin symmetry and its breaking in nuclear physics. The isospin symmetry is an approximate symmetry and is broken
explicitly by Coulomb and isospin symmetry breaking (ISB) terms in nucleon-nucleon interaction. In this work, we shall investigate the
implications of both of these on the nuclear properties. Further, isospin symmetry is also broken spontaneously due to the mean-field
approximation which shall be also discussed in this work. 
The review article is organised in the following manner: Basic ideas regarding
the isospin symmetry are presented in  
Sec.~\ref{sec:iso_bas}. The quantum chromodynamics (QCD) origin of the ISB
is discussed in Sec.~\ref{sec:iso_QCDSCSB}. The classification of charge-dependent interactions between nucleons are presented in Sec.~\ref{sec:iso_ccdnn}. Sec.~\ref{sec:iso_exp} discusses the empirical data, manifesting the
breakdown of the isospin symmetry in hadronic systems. In Sec.~\ref{sec:iso_IMME}, the important consequences of the isospin symmetry in nuclear physics, i.e., isobaric mass multiplet equation (IMME) and Nolen-Schiffer anomaly are presented. Sec.~\ref{sec:iso_dft} presents the detailed investigation
of the ISB using the density functional approach.
The spherical shell model study of the ISB is discussed in Sec.~\ref{sec:iso_ssm}.
Summary and concluding remarks are included in Sec.~\ref{sec:iso_summary}.

\section{ Basic concepts of isospin symmetry }
\label{sec:iso_bas}

It was in the year 1932 when the neutron was discovered \cite{Chadwick1932} by Chadwick and Heisenberg
realised \cite{Heisenberg1989}  that neutron and proton
could be treated as two states of the same particle called "nucleon", considering that the two particles have similar
masses and the same intrinsic spin.
The two states of the nucleon are distinguished by the values of a new quantum number that was called as "isospin".
The name isospin was christened by Wigner
in 1937 \cite{Ewigner1937} to express the similarity between
this quantity and the spin quantum number. Although the mathematical structure of the two quantities is identical, but
in comparison to the spin which is an intrinsic property of a particle, the isospin symmetry is invoked. Nevertheless,
the concept of isospin has played a fundamental role in the development of quark model in particle physics \cite{Gellmann1964}.
Further, in 1959 Weinberg derived the isospin selection rule for particle decay,
$\Delta T \le 2$, by recognising that Coulomb interaction weakly breaks the isospin symmetry \cite{Weinberg1959}.

In nuclear physics, approximate models have been developed using the concept of isospin symmetry. For instance,
the famous isobaric mass multiplet equation (IMME) of Wigner \cite{Ewigner1957} to predict the masses of isobaric chains of isotopes has its roots
in the isospin quantum number. In Sec.~\ref{sec:iso_IMME}, we shall discuss in detail the importance of this
mass formula in predicting the masses of members of a multiplet which are difficult to measure experimentally. The nucleon-nucleon
interaction in most of the approaches is also expanded in terms of the isospin quantum number as this allows to evaluate the
effects of the ISB terms \cite{Henleybook1969}.

In analogy to the spin quantum number of an electron with $s = \frac {1} {2}$, the
isospin quantum number of the nucleon is defined as, $t = \frac {1} {2}$ with $t_z = +1/2 (-1/2)$ for a neutron (proton).
The total isospin projection for a nuclear system, having a total "$A$" number of nucleons, is given by
\begin{equation}
  T_z = \sum_{i=1}^{A} t_z(i) = \frac{(N-Z)}{2} ~,
\end{equation}
with the possible range of isospin quantum number, $T \ge |T_z|$. The projection quantum number is well defined for
a given system, but isospin itself is not conserved in general. For conservation of the isospin quantum number, it is
required that the isospin operator must commute with the Hamiltonian of the system \cite{Miller1990}, i.e.,
\begin{eqnarray}
 [\widehat{H},{\widehat{\bf{T}}}]=0~~~~~~\Rightarrow~~~~~~[\widehat{H},\widehat{\bf{T}}^2]=0~,
 \label{HT1}
\end{eqnarray}
 where the isospin operator can be written in terms of Pauli matrices
 \begin{eqnarray}
   {\widehat{\bf{T}}}=\frac{1}{2} {\boldsymbol{\tau}}~,
\end{eqnarray}
with 
 \begin{eqnarray}
   {\boldsymbol{\tau}}=\hat{i}~{\tau}_x+ \hat{j}~{\tau}_y+\hat{k}~{\tau}_z~,
 \end{eqnarray}
 \begin{eqnarray}
   {\tau}_x=\left( \begin{array}{cc}
                  0 & 1 \\
                  1 & 0 
                 \end{array}\right)~,
 \qquad 
   {\tau}_y=\left( \begin{array}{cc}
                  0 & -i \\
                  i & 0 
                 \end{array}\right)~,
 \qquad  
   {\tau}_z=\left( \begin{array}{cc}
                 1 & 0 \\
                  0 & -1 
                \end{array}\right)~.
 \end{eqnarray}
 The simplest possible nucleon-nucleon interaction that satisfies the commutation relation of Eq. (\ref{HT1}) is of scalar form in the isospace, i.e., 
 \begin{eqnarray}
  \widehat{V}_{ij}=a+b\, {\boldsymbol{\tau}}(i)\cdot{\boldsymbol{\tau}}(j)~,
 \end{eqnarray}
where the coefficients "$a$" and "$b$" depend on space and spin operators.
The isospin invariance of a system implies that its Hamiltonian commutes with the rotation operator about any axis defined in the
isospace. This is commonly referred to as charge independence (CI) in the
literature \cite{Miller1990,Bentley2007}. The
charge symmetry (CS) operator, another related quantity defined in nuclear physics, indicates a rotation by $\pi$ about
an axis perpendicular to the quantization axis along which the isospin projection quantum number is measured.
Conventionally, z-axis is considered as the quantization axis, and y-axis is chosen as the rotational axis. 
The CS operator is then given by
 \begin{equation}
  \widehat{P}_{CS}=\textrm{exp}(i\pi \widehat{T}_y)~,~~~~ \mbox{with} \qquad\widehat{T}_y=\sum_{i=1}^A\widehat{t}_y(i)=\frac{1}{2}\tau_y~.
  \label{PCS}
 \end{equation}
If CS is conserved then the corresponding CS operator, $ \widehat{P}_{CS}$ given by Eq.~{\ref{PCS}}, commutes with the Hamiltonian.
The above rotation operator will interchange "$z$" and "$-z$" axis in the isospace, which means
that $T_z$ of a state will change into
$-T_z$. Therefore, CS symmetry implies the equivalence of neutron-neutron (nn) and proton-proton (pp) forces.
It has been demonstrated in Ref. \cite{Miller1990} that this equality is a necessary condition, but not 
sufficient for the validity of CS. The symmetry also has implications for the neutron-proton system and is shown,
following the derivation given in Ref. \cite{Miller1990}, by considering the commutator of
$ \widehat{T}_z$ with the Hamiltonian operator acting on a self-conjugate state $|T_z=0\rangle$
 \begin{align}
   [\widehat{P}_{CS},\widehat{T}_z]|T_z=0\rangle&=(\widehat{P}_{CS}\widehat{T}_z+\widehat{P}_{CS}\widehat{T}_z)|T_z=0\rangle~,\nonumber\\
   &=2\widehat{P}_{CS}\widehat{T}_z|T_z=0\rangle~,\nonumber\\
   &=0~,
   \label{PcsH3}
 \end{align}
where we have made use of the following relation along with the fact that $ \widehat{P}_{CS}$ is a unitary operator, i.e., $\widehat{P}_{CS}\widehat{P}_{CS}^\dagger=I$,
 \begin{eqnarray}
   \widehat{P}_{CS}^\dagger \widehat{T}_z\widehat{P}_{CS}=-\widehat{T}_z\qquad\qquad \rm or\qquad\qquad
    \widehat{T}_z\widehat{P}_{CS}=-\widehat{P}_{CS}\widehat{T}_z~.\nonumber
   \label{PcsH}
 \end{eqnarray}
 If $\widehat{P}_{CS}$ commutes with the Hamiltonian for the self-conjugate system, then we can specify the
 charge (eigenvalue of $T_z$)
 and a charge parity quantum number, which is the eigenvalue of $\widehat{P}_{CS}$, simultaneously. Since the application of $\widehat{P}_{CS}^2$ brings the system back to its original state, i.e.,
 \begin{eqnarray}
  \widehat{P}_{CS}^2&=&I~,
 \end{eqnarray}
the eigenvalues of $\widehat{P}_{CS}$ are $\pm 1$. For a state with a well defined value of isospin, we obtain
 \begin{align}
   \widehat {P}_{CS}|\alpha, T, T_z=0\rangle &= \textrm{exp}(i\pi \widehat{T}_y)|\alpha, T, T_z=0\rangle~,\nonumber\\
   &=\sum_{T_z^\prime}\langle \alpha, T, T_z^\prime|\textrm{exp}(i\pi
   \widehat{T}_y)|\alpha, T, T_z=0\rangle |\alpha, T, T_z^\prime\rangle~,\nonumber\\
  &=\sum_{T_z^\prime}d_{T_z^\prime 0}^T(\pi)|\alpha, T, T_z^\prime\rangle~,\nonumber\\
  &=(-1)^{T}|\alpha, T, T_z=0\rangle~.
  \label{dalpha}
 \end{align}
 The above equation gives the eigenvalues of CS operator in terms of the isospin quantum number, implying isospin conservation
 for self-conjugate nuclei obeying charge symmetry.

 
 It can be shown that CS operator $\widehat{P}_{CS}$ converts a proton into a neutron or vice-versa
only when the bare masses are considered and for physical masses a perturbative
correction term needs to be considered \cite{Miller1990}. The term "bare" is the eigenstate of the Hamiltonian $\widehat{H}_0$ in the
absence of electromagnetic interactions and other charge breaking terms. Denoting the "bare" neutron (proton) of momentum ${\bf k}$, and magnetic quantum number
$m_s$ by $|n(p){\bf k}m_s)$, such that
\begin{eqnarray}
  |p{\bf k}m_s)= \left( \begin{array}{c}
    0\\ 1
  \end{array}\right) |{\bf k}m_s)~,
  \label{pkms1}
\end{eqnarray}
and
 \begin{eqnarray}
  |n{\bf k}m_s)= \left( \begin{array}{c}
    1\\ 0
\end{array}\right) |{\bf k}m_s)~,
 \end{eqnarray}
 where the column vectors denote the isospin part of the 
 wave function. It is noted that momentum and spin remain unaffected due to $\widehat{P}_{CS}$ operator. Therefore, we have
 \begin{align}
  \widehat{P}_{CS}|p{\bf k}m_s)= \textrm{e}^{i (\pi/2)\tau_y}\left(\begin{array}{c}
                  0 \\
                  1 
   \end{array}\right) |{\bf k}m_s)& =\left[1+\frac{i\pi}{2}\tau_y+\frac{1}{2!}(\frac{i\pi}{2})^2(\tau_y)^2+\frac{1}{3!}
   (\frac{i\pi}{2})^3(\tau_y)^3\right.\nonumber\\
   &~~~~+\left.\frac{1}{4!}(\frac{i\pi}{2})^4(\tau_y)^4+...\right]\left( \begin{array}{c}
                 0 \\
                  1 
   \end{array}\right) |{\bf k}m_s)\nonumber\\ 
   &={\cos}\frac{\pi}{2}\left( \begin{array}{c}
                  0\\
                  1 
   \end{array}\right) |{\bf k}m_s)
   +{\sin}\frac{\pi}{2}\left( \begin{array}{c}
                  1 \\
                  0 
   \end{array}\right) |{\bf k}m_s)\nonumber\\
   &=\left( \begin{array}{c}
                  1 \\
                  0 
   \end{array}\right) |{\bf k}m_s)\nonumber\\
   &=|n{\bf k}m_s)~,
   \label{eoperator}
 \end{align}
 where the properties of the Pauli matrices have been used.
 Therefore, we obtain
 \begin{eqnarray}
   \widehat{P}_{CS}|p{\bf k}m_s)=|n{\bf k}m_s)~.
   \label{PCSN}
 \end{eqnarray}
In a similar manner, we can show that
 \begin{eqnarray}
   \widehat{P}_{CS}|n{\bf k}m_s)=-|p{\bf k}m_s)~.
   \label{PCSP}
 \end{eqnarray}
  In the relativistic version,
 $\widehat{P}_{CS}$ in the helicity representation, and is given by \cite{Miller1990}
 \begin{eqnarray}
  \widehat{P}_{CS}|p{\bf k}\lambda)=|n{\bf k}\lambda)~,~~~~~~~~~~~~~~~~~~  \widehat{P}_{CS}|n{\bf k}\lambda)=-|p{\bf k}\lambda)~.
   \label{PCSP1}
 \end{eqnarray}
As indicated above, for the "bare" neutron-proton system, we have $[\widehat{H}_0, \widehat{P}_{CS}]=0$. The corresponding eigen value equation is given by
 \begin{eqnarray}
   \widehat{H}_0|n(p){\bf k}\lambda)=\sqrt{k^2+M_0^2}~~~|n(p){\bf k}\lambda)~,
   \label{H0}
 \end{eqnarray}
 where $M_0$ is the "bare" nucleon mass. In the presence of the CSB terms, which are considered as the perturbation Hamiltonian $\widehat{H}_1$ to the charge symmetric Hamiltonian, $\widehat{H}_0$, such that
 the total Hamiltonian resulting the physical neutron and proton states is given by
 \begin{eqnarray}
 \widehat{H}= \widehat{H}_0+ \widehat{H}_1~.
 \end{eqnarray}
 The eigen value equations of this total Hamiltonian are given by
 \begin{eqnarray}
   \widehat{H}|n{\bf k}m_s\rangle=\sqrt{k^2+M_n^2}~~~|n{\bf k}m_s\rangle~,
   \label{HNP1}
 \end{eqnarray}
 and
 \begin{eqnarray}
   \widehat{H}|p{\bf k}m_s\rangle=\sqrt{k^2+M_p^2}~~~|p{\bf k}m_s\rangle~.
   \label{HNP2}
 \end{eqnarray}
 Thus, the  "bare" strong Hamiltonian has two-fold degeneracy, resulting from its charge symmetry. This symmetry is broken by the CSB terms resulting into the physical neutron and proton with masses given by  $M_n=939.6$MeV and $M_p=938.3$MeV. In the above equations, the complete physical states are distinguished by sharp kets $|~~~\rangle$ instead of the round one $|~~~)$ of Eqs. (\ref{pkms1}) to (\ref{H0}) for the "bare" states.  
 Considering $\widehat{H}_1$ as a perturbative correction, it is shown in appendix (\ref{appendix1}), how the physical states can be obtained in the lowest order of the perturbation theory.

 \section{Quantum Chromodynamics and charge symmetry breaking}
 \label{sec:iso_QCDSCSB}

 The microscopic origin of the CSB is the electromagnetic interaction and the mass difference
 between up- and down-quarks. Therefore, the investigation of CSB in nuclear physics will provide an insight into the
 quark mass problem and may also give some clues on the flavor dependence of the quark mass spectrum, which
 is one of the most fundamental unsolved problems in physics \cite{Machleidt_2016}. In the following, it
 is demonstrated how mass difference
 between up- and down-quarks leads to CSB mechanism. We start with the QCD Lagrangian \cite{Machleidt_2016} $:$ 
\begin{eqnarray}  
{\mathcal{L}}_{\textrm{QCD}}=\overline{q}(i\gamma ^\mu \mathcal{D}_\mu-\mathcal{M})q-\frac{1}{4}\mathcal{G}_{\mu\nu,a}\mathcal{G}_a^{\mu\nu}~,
 \label{LQCD}
\end{eqnarray}
with the covariant derivative
\begin{eqnarray}
  \mathcal{D}_{\mu}=\partial_\mu-ig\frac{\lambda_a}{2}\mathcal{A}_\mu,a~,
\label{DMU}  
\end{eqnarray}
and the gluon field tensor
\begin{eqnarray}
  \mathcal{G}_{\mu\nu,a}=\partial_\mu \mathcal{A}_{\nu,a}-\partial_\nu \mathcal{A}_{\mu,a}+gf_{abc}\mathcal{A}_{\mu, b}\mathcal{A}_{\nu, c}~.
\label{GMU}  
\end{eqnarray}
In above Eqs. ~(\ref{LQCD}) to (\ref{GMU}), "$q$" denote the quark fields, $\mathcal{M}$ the quark mass matrix, $g$ is the strong coupling constant
and $\mathcal{A}_{\mu, a}$ are the gluon fields. $\lambda_a$ are the Gell-Mann matrices and $f_{abc}$ the structure
constant of the $SU(3)_{\textrm{color}}$ Lie algebra with $a,b,c=1,2,...,8$.
The first generation of quark masses at 2 GeV scale in the minimal subtraction (MS) scheme from $(2+1)$ flavour lattice QCD with
estimated quantum electrodynamics (QED) correction are given by \cite{Tetsuo2012}
\begin{align*}
  m_u&=2.19 (15)~\textrm{MeV}~,\\
  m_d&=4.67(20)~\textrm{MeV}~,\\
  m_s&=94(3)~\textrm{MeV}~.
\end{align*}
In the absence of the quark masses, the QCD Lagrangian is given by
\begin{eqnarray}
  \mathcal{L}_{\textrm{QCD}}^0=\overline{q}i\gamma ^\mu \mathcal{D}_\mu q-\frac{1}{4}\mathcal{G}_{\mu\nu,a}\mathcal{G}_a^{\mu\nu}~.
  \label{LQCD1}
\end{eqnarray}
Defining right- and left-handed quark fields as
\begin{eqnarray}
  q_R=P_{R}~q~,~~~~~~~q_L=P_{L}~q~,\nonumber
\end{eqnarray}
where
\begin{eqnarray}
  P_R=\frac{1}{2}(1+\gamma_5)~,~~~~~~~P_L=\frac{1}{2}(1-\gamma_5)~.
\end{eqnarray}
Lagrangian of Eq. (\ref{LQCD}) can then be rewritten as
\begin{eqnarray}
\mathcal{L}_{\textrm{QCD}}^0=\overline{q}_Ri\gamma ^\mu \mathcal{D}_\mu q_R+\overline{q}_Li\gamma ^\mu \mathcal{D}_\mu q_L-\frac{1}{4}\mathcal{G}_{\mu\nu,a}\mathcal{G}_a^{\mu\nu}~.
\end{eqnarray}
It is evident from the above equation that left- and right-handed components of the massless quarks do not mix. In the two-flavor case, it is $SU(2)_R \times SU(2)_L$ symmetry, what is called as chiral symmetry, that is preserved.
It is shown below that this symmetry is broken by considering the mass term 
\begin{align}
\mathcal{M}&=\left( \begin{array}{cc}
   m_u & 0\\
   0 & m_d
\end{array}\right)\nonumber\\
&=\frac{1}{2}(m_u+m_d)\left( \begin{array}{cc}
   1 & 0\\
   0 & 1
\end{array}\right)
+\frac{1}{2}(m_u-m_d)\left( \begin{array}{cc}
   1 & 0\\
   0 & -1
\end{array}\right)\nonumber\\
&=\frac{1}{2}(m_u+m_d)I+\frac{1}{2}(m_u-m_d)\tau_3~.
\end{align}
In the above equation, the chiral symmetry is broken even when
the two masses are considered to be same. On the other hand, the first term conserves isospin symmetry
but the second term breaks it. For $m_u=m_d$, the isospin symmetry is preserved by the QCD Lagrangian. Therefore, it is evident that mass difference between up- and down-quarks leads to the ISB, apart from the electromagnetic interactions.

\section{ Classification of charge-dependent nucleon-nucleon interaction}
\label{sec:iso_ccdnn}
Following the Henley and Miller classification scheme \cite{Henleybook1969,Henleybook1979,Miller1990}, nucleon-nucleon
interaction is categorised in terms of the following groups:
\begin{enumerate}
\item {\bf Class I}~: ~Isospin or charge-independent interaction commutes with the isospin operator
\begin{eqnarray}
  [\widehat{V}_I, \widehat{{\bf T}}]=0~,
\end{eqnarray}
and can be written as a scalar operator in the isospace
\begin{eqnarray}
  \widehat{V}_I=a+b~{\bm \tau}(i).{\bm \tau}(j)~,
\end{eqnarray}
where "$a$" and "$b$" are isospin independent quantities.
\item {\bf Class II}~:~ The interaction in this category is charge symmetric, but breaks charge independence and can be written as
\begin{eqnarray}
  \widehat{V}_{II}=c~[\tau_3(i)\tau_3(j)-\frac{1}{3}{\bm\tau}(i).{\bm\tau}(j)]~.
  \label{VII}
\end{eqnarray}
In the above equation, the charge symmetry operator, $e^{-\frac{i\pi}{2}\tau_y}$, changes sign for $\tau_3(j)$ and again for $\tau_3(i)$ and therefore first term is invariant under the charge symmetry transformation. However, under an arbitrary rotation in the isospace, this term is not invariant and therefore breaks the charge independence. The second term in the above equation is isoscalar. Coulomb interaction has the form of Eq. (\ref{VII}).
\item {\bf Class III}~:~ This class of forces break both charge symmetry and charge independence, but are symmetric under exchange of particles
  in isospace and is written as
\begin{eqnarray}
  \widehat{V}_{III}=d~[\tau_3(i)+\tau_3(j)]~,
\end{eqnarray}
where "$d$" is symmetric under interchange of spin coordinates. This above class of force is different for protons and neutrons, and vanishes for the neutron-proton system. However, it does not lead to isospin mixing as $[\widehat{V}_{III}, \widehat{T}_{z}]=0$ and  $[\widehat{T}^2, \widehat{T}_{z}]=0$. The $\rho^0-\omega$ meson mixing gives significant contribution to this class of force and also the neutron-proton mass difference in two-pion-exchange potential.
\item {\bf Class IV}~:~ This class of forces break charge symmetry and consequently the charge independence and is generally written as
\begin{eqnarray}
  \widehat{V}_{IV}=e\biggl[\biggl({\bm\sigma}(i)-{\bm\sigma}(j)\biggl).{\bf L}\biggl]\biggl[\tau_3(i)-\tau_3(j)\biggl]+f\biggl[\biggl({\bm\sigma}(i)\times {\bm\sigma}(j)\biggl).{\bf L}\biggl]\biggl[{\bm\tau}(i)\times {\bm\tau}(j)\biggl]_3~.
\end{eqnarray}
The main contribution to the first term comes from photon and $\rho-\omega$ exchange terms, and for the second term,
it is from the nucleon mass difference on $\pi$ and $\rho$ exchanges. It vanishes for neutron-neutron and proton-proton systems, but causes
isospin-mixing in the neutron-proton system.
  \end{enumerate}

\section{ Empirical data on isospin symmetry breaking }
\label{sec:iso_exp}

In this section, we shall review the empirical evidence on the breaking of the isospin symmetry in nuclear physics.

\subsection{Mass differences in mirror and analogue states}

\begin{table}[H]
  \caption{The multiplet mass splitting of the lightest nuclear isodoublets.
    \label{quark}}
\newcolumntype{C}{>{\centering\arraybackslash}X}  
\begin{tabularx}{\textwidth}{|C|C|C|C|}
\hline	
{\bf Nucleus}	&   ${\bf T},~{\bf J}^{\pi}$		&{\bf Mass}	& {\bf Mass difference}	\\
	&		&{\bf (MeV)}	& {\bf (MeV)}	\\
\hline				
					
n	&$\frac{1}{2}$, $\frac{1}{2}^+$		&939.57	&$+1.29$	\\
p	&	&938.28	&	\\\hdashline
$^3$H	&$\frac{1}{2}$, $\frac{1}{2}^+$		&2808.94	&$+0.52$	\\
$^3$He	&		&2808.42	&	\\\hdashline
$^5$He	&$\frac{1}{2}$, $\frac{3}{2}^-$		&4667.87	&$+0.21$	\\
$^5$Li	&		&4667.66	&	\\\hdashline
$^7$Li	&$\frac{1}{2}$, $\frac{3}{2}^-$		&6533.89	&$-0.35$	\\
$^7$Be	&		&6534.24	&	\\
\hline
\end{tabularx}
\end{table}
\begin{figure}[htp!]
 \centerline{\includegraphics[trim=0cm 0cm 0cm
0cm,width=1\textwidth, height=1.3\textwidth, clip]{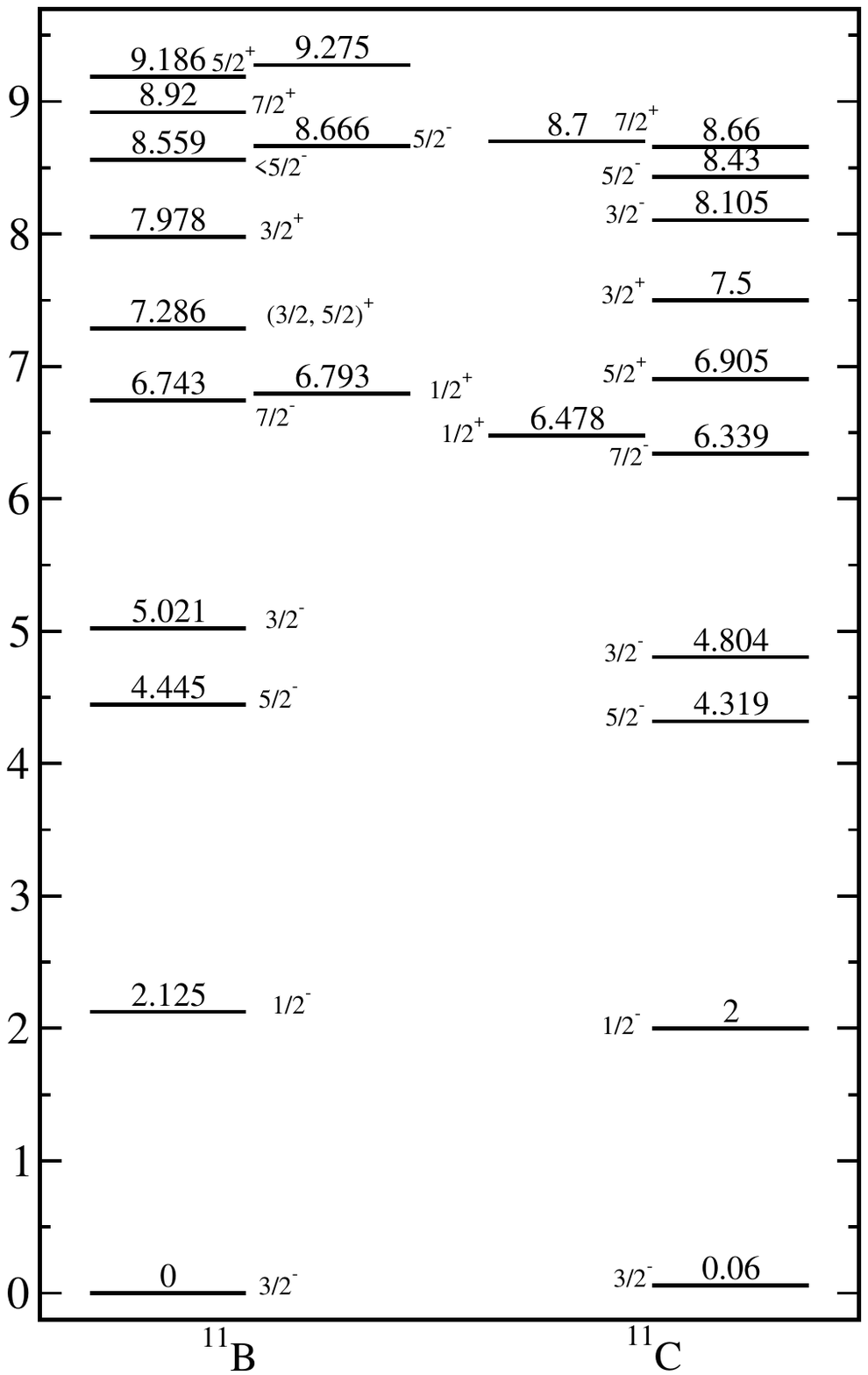}} \caption{(Color
online) Approximate charge symmetry of nuclear levels in mirror nuclei.
  }
\label{CS1}
\end{figure}
\begin{figure}[htb]
 \centerline{\includegraphics[trim=0cm 0cm 0cm
0cm,width=0.7\textwidth,height=0.8\textwidth,clip]{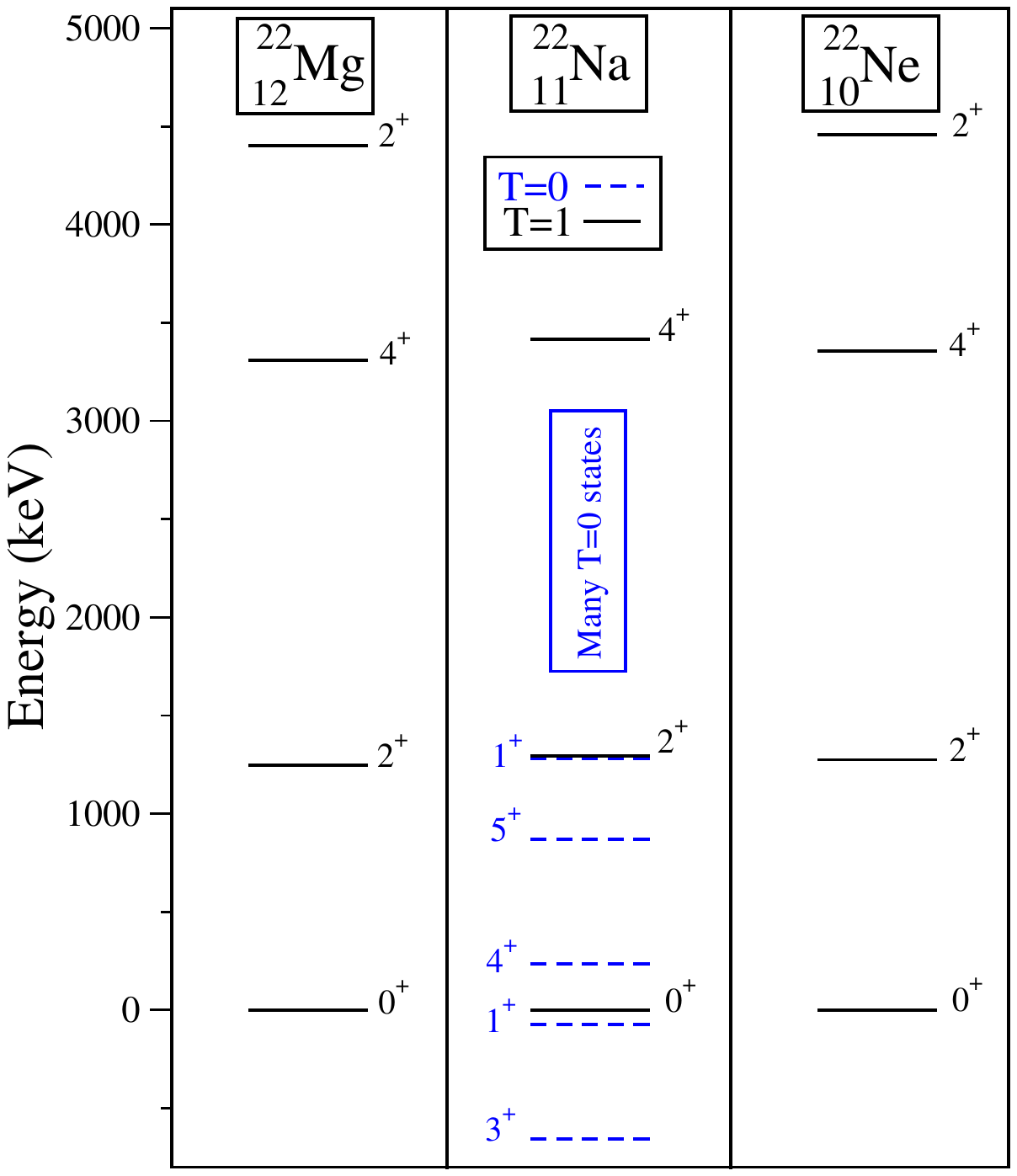}} \caption{(Color
online) The spectra of $A=22$ triplets. 
 }
\label{Figisobaric}
\end{figure}

It has been discussed in the introduction section that the idea of the isospin symmetry was conceived when neutron was discovered
and Heisenberg \cite{Heisenberg1989} realised that since mass of this new particle is similar to that of a proton, the two particles can be treated on equal
footing. The mass difference between proton and
neutron, a fundamental constant in physics, is 1.29 MeV. This mass difference is a direct manifestation of the ISB and in fact is very critical for the very existence of the universe
\cite{wilczek2015}. In the event that this mass difference was less than that of electron mass of 0.51 MeV, the proton will capture
electrons from the atomic shells through the process called as inverse $\beta$-decay with the consequence that an atom will cease to
exist. In the other scenario that this difference is
larger than the observed one, it will be impossible to synthesise the elements beyond hydrogen \cite{wilczek2015}.

The ISB terms of electromagnetic origin and u-d mass splitting give rise to the neutron-proton mass difference. In a 
major achievement \cite{Borsanyi2015}, QED and QCD equations have been simultaneously solved
on a lattice using powerful computers and the masses of hadrons have been determined using four non-degenerate Wilson fermion flavors.
In particular, the neutron-proton mass difference
has been accurately reproduced by the calculations and it has shown that it results from the competition between electromagnetic
and u-d mass ISB terms. Although the u-d mass difference is a parameter in the model as this quantity is not
known empirically, however, using the same u-d mass difference, the masses of a large set of particles having different quark structures
have been evaluated and are found to be consistent with the measured values \cite{Borsanyi2015}.

In Table \ref{quark}, the masses of light mirror nuclei are listed. The isospin symmetry signifies that the masses
of the two mirror nuclei (isospin doublets) should be identical and any difference shall be attributed to ISB terms in the Hamiltonian. The major contributor to the ISB is
the Coulomb interaction as one mirror partner has more protons than the other and will have larger Coulomb energy. We shall discuss later
that nuclear interaction also breaks the isospin symmetry and contributes to the mass difference.
Isospin symmetry, as a matter of fact, entails that not just masses but all the properties of the mirror nuclei should be identical \cite{Wachter1988} apart
from the corrections arising from the Coulomb interaction. In Fig.~(\ref{CS1}), the energy spectrum of the mirror nuclei, $^{11}$B and
$^{11}$C are displayed, and the figure depicts a remarkable similarity
between the two spectra, a consequence of the isospin symmetry.

\begin{figure}[H]
 \centerline{\includegraphics[trim=0cm 0cm 0cm
0cm,width=0.6\textwidth,height=0.4\textwidth,clip]{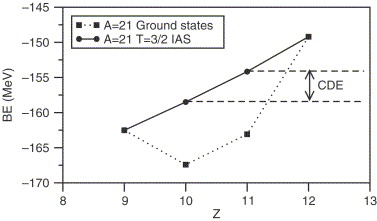}} \caption{(Color
online) The experimental binding energies of the $A=21, T_z=\pm \frac{1}{2}, \pm \frac{3}{2}$ nuclei - both for ground states and for the $T=\frac{3}{2}$ isobaric analogue states. The Coulomb displacement energy (CDE) between adjacent members of the $T=\frac{3}{2}$ quadruplet is indicated. (Adopted from Ref. \cite{Bentley2007}).}
\label{Fig1111}
\end{figure}

The above similarity is not confined to mirror nuclei only, but holds for a broader class of states, so called,
the isobaric analogue states. For a given angular-momentum, a state of a nucleus with $T$ greater than $T_z$ is called 
an analogue state as it has identical structure to the state of a neighbouring isobar with $|T_z|=T$. An example of
isobaric analogue states of $A=22$ triplet are depicted in Fig.~(\ref{Figisobaric}). $^{22}$Mg and $^{22}$Ne having two-protons and two-neutrons
as valence particles have $T=1$ states only, whereas $^{22}$Na having neutron-proton pair has both $T=1$ and 0 states. It is noted from the
figure that after shifting the spectrum of $^{22}$Na downwards to match the lowest $T=1$ ($J^\pi=0^+$) states of the three nuclei, the other
$T=1$ states with $J^\pi=2^+, 4^+$ and $6^+$ fall in place. The difference in the energies of these states or any other properties is the indication
of charge independence breaking (CIB). To further illustrate the the meaning of analogue states, in Fig.~(\ref{Fig1111}) the binding of the
$A=21$ $T=3/2$ quadruplet states are depicted.  For the  two extreme cases, the ground-states are having $T=3/2$ and for the two
middle cases, the first excited states belong to the $T=3/2$ quadruplet. The ground-states for the latter two cases are having $T=1/2$.
For mirror nuclei, only proton and neutron pairs are exchanged  and it is possible to test CSB only. However,
for isobaric analogue states of three or more nuclei, CIB is investigated.

\subsection{Scattering lengths}
The $^{1}S_0$ scattering length provides a direct evidence of CSB and CIB effects in the nuclear
force \cite{Miller2006}.
In Table \ref{slength}, neutron-neutron, proton-proton and neutron-proton singlet scattering length parameters are listed after correcting for the
electromagnetic effects. 
To measure CIB effects, one defines the quantity, $\Delta a_{CD}$, as
\begin{equation}
\Delta a_{CIB} = \frac {1} {2} \biggr(a^N_{pp} + a^N_{nn}\biggr) - a^N_{np} = 5.7 \pm 0.3 {\rm fm}~,
\end{equation}
and for CSB the quantity, $\Delta_{CSB}$, is defined as
\begin{equation}
\Delta a_{CSB} = a^N_{pp} - a^N_{nn} = 1.5 \pm 0.5 {\rm fm}~.
\end{equation}
\begin{table}[H]
\caption{Charge independence of nuclear Hamiltonian imposes the equalities of the NN scattering lengths; $a^N_{pp}$=$a^N_{nn}$=$a^N_{np}$ . However, the experimentally measured $^{1}S_0$ scattering lengths differ, a consequence of CIB of the Hamiltonian. \label{slength}}
\newcolumntype{C}{>{\centering\arraybackslash}X}
\begin{tabularx}{\textwidth}{|C|C|}
\hline	
	{\bf Nucleon pair}& {\bf Scattering length (fm)}\\
\hline									
$a^N_{pp}$&-17.3$\pm$0.4\\
$a^N_{nn}$&-18.8$\pm$0.3\\
$a^N_{np}$&-23.77$\pm$0.09\\				
\hline
\end{tabularx}
\end{table}
The above differences in the scattering lengths translate into  $2.5\%$ and
$0.6\%$ changes in the potential \cite{Miller2006}. These differences are significant and need to be considered to describe the nuclear
processes accurately. 
The present level of understanding is that there are primarily two mechanisms behind the measured ISB effects in nuclear force.
The first is the mass splitting observed for hadrons in the isospin channel, and the second is the meson mixing and the irreducible
meson-photon exchanges. This leads to a smaller kinetic energy for neutrons as compared to protons and increases the neutron-neutron
scattering length by 0.25 fm in comparison to the proton-proton one. Besides this, the nucleon mass asymmetry also contributes to the
meson exchange diagrams of $2 \pi$ and $ N \Delta$ types that contribute to the nuclear force \cite{Miller1990}.

The up- and down-quark mass difference and the electromagnetic interaction also leads to mixing of neutral mesons with same spin and
parity, but different isospin \cite{Miller1990}. The most important mixing is between $\rho^0$ and $\omega$ neutral mesons and it has been
shown \cite{Miller1990} that this mixing can completely explain the observed CSB in the singlet state. However, it has been remarked that
only on-shell mixing matrix elements have been extracted, and the calculations are not quite accurate \cite{Miller1990}.

In an attempt to unravel the source of CSB in nuclear force, three different models have been studied \cite{Machleidt2001}. These models include,
$\rho-\omega$ mixing, nucleon mass splitting and the phenomenological Argonne potential ($V_{18}$). It has been shown that all the three
models give similar values for the $^{1}S_0$ scattering length. Although considerable departures were noted in these models for ISB
in $^{3}P_J$ partial waves, but the contribution of these to some observable properties, like the binding energy of the mirror
pair $^{3}$H-$^{3}$He were found to be of the order of only $6\%$. Further studies are being planned to probe the cause of the
ISB effects in nuclear force \cite{Machleidt2001}. 

\section{ Isobaric mass multiplet equation (IMME) and Nolen-Schiffer anomaly}
\label{sec:iso_IMME}
In this section, the IMME and the Nolen-Schiffer anomaly are discussed.
\subsection{Isobaric mass multiplet equation}
In this section, we shall discuss the IMME, which is one of the most important consequences of the
isospin symmetry in nuclear physics \cite{Bentley2007}. It was demonstrated by Wigner \cite{Ewigner1957} that the mass difference of the
isospin analogue states can be expressed in terms of a quadratic polynomial in $T_z$. This was shown using the first order
perturbation correction arising from the Coulomb interaction, 
which explicitly breaks the isospin symmetry in nuclei. The correction
was evaluated using the basis functions having well defined isospin quantum number. To begin with, it may be noted that the Coulomb interaction has the following tensorial 
decomposition in the isospace :

\begin{align}
  \widehat{H}_c&=e^2\sum_{i<j}\frac{(1/2-\widehat{t}_{z}(i))(1/2-\widehat{t}_{z}(j))}{r_{ij}} ~,\nonumber\\
  &=e^2\sum_{i<j}\frac{1}{r_{ij}}\left(\frac{1}{4}-\frac{1}{2}(\widehat{t}_z(i)+\widehat{t}_z(j))+\widehat{t}_z(i)\widehat{t}_z(j)\right) ~,\nonumber\\
  &=e^2\sum_{i<j}\frac{1}{r_{ij}}\biggl[\biggl(\frac{1}{4}+\frac{\widehat{\bf t}(i)\cdot \widehat{\bf t}(j)}{3}\biggl)
    -\biggl(\frac{\widehat{t}_{z}(i)+\widehat{t}_{z}(j)}{2}\biggl)+  \biggl(\widehat{t}_{z}(i)\widehat{t}_{z}(j)-\frac{\widehat{\bf t}(i)\cdot\widehat{\bf t}(j)}{3}\biggl)\biggl]~.
  \label{HCE}
\end{align}
The first, second and third terms in Eq.~(\ref{HCE}) are isoscalar, isovector and isotensor components of the Coulomb interaction,
respectively. In an analogous manner, the total Hamiltonian of the nuclear many-body system is considered to have the following
isospin expansion in terms of $T=0$ and 1 channels \cite{ORMAND19891}~~:
\begin{eqnarray}
  \widehat{H}_{\textrm{tot}}=\widehat{v}_0^{(0)}\widehat{I}_0^{(0)}+\sum_{k=0}^2\widehat{v}_1^{(k)}\widehat{I}_1^{(k)}~,
  \label{HTOT1}
\end{eqnarray}
where the operator $\widehat{I}_{T}^{(k)}$ separates the explicit isospin dependence from the radial and spin components of the nucleon-nucleon interaction and are given by \cite{ORMAND19891}
\begin{eqnarray}
  \widehat{I}_0^{(0)}&=&\frac{1}{4}-\widehat{\bf t}(1).\widehat{\bf t}(2)~,~~~~\widehat{I}_1^{(0)}=\frac{3}{4}+\widehat{\bf t}(1).\widehat{\bf t}(2)~,\nonumber\\
  \widehat{I}_1^{(1)}&=&\frac{1}{2}\biggl(\widehat{t}_z(1)+\widehat{t}_z(2)\biggl)~,~~~~\widehat{I}_1^{(2)}=\widehat{t}_z(1)\widehat{t}_z(2)-\frac{1}{3}\widehat{\bf t}(1).\widehat{\bf t}(2)~.
\end{eqnarray}
In Eq.~(\ref{HTOT1}), the $T=1$ component of nucleon-nucleon interaction, involving the radial and the spin components can be written as
\begin{equation}
  v_1^{(k)} = \sum_{\mu} S^{(k)}_\mu V_\mu(r)~,
  \label{V1k}
\end{equation}
where the summation over "$\mu$" includes Coulomb, Yukawa and the isoscaler terms of the nucleon-nucleon interaction, and the quantities $S$ and $V(r)$
represent the spin and radial dependence of the interaction terms. In the following, we shall express the interaction in terms of
neutron-proton, neutron-neutron and proton-proton terms, which allows to delineate the terms in Eq.~(\ref{V1k}) 
corresponding to ISB and CSB mechanisms. As an illustration, we consider the evaluation of the following matrix element~~:


\begin{align}
  \biggl\langle p(1)p(2)&\biggl|(\widehat{\bf t}(1).\widehat{\bf t}(2))\biggl|p(1)p(2)\biggl\rangle\nonumber\\
  &=\sum_TC\biggl(\frac{1}{2}\frac{1}{2}T;-\frac{1}{2}-\frac{1}{2}-1\biggl)C\biggl(\frac{1}{2}\frac{1}{2}T;-\frac{1}{2}-\frac{1}{2}-1\biggl)\nonumber\\
  &~~~~~~~~~~~~~~~~~~~\times~~ \prescript{}{T}{\biggl\langle} p(1)p(2)\biggl|(\widehat{\bf t}(1).\widehat{\bf t}(2))\biggl|p(1)p(2)\biggl\rangle_{T}\nonumber\\
  &=C\biggl(\frac{1}{2}\frac{1}{2}1;-\frac{1}{2}-\frac{1}{2}-1\biggl)C\biggl(\frac{1}{2}\frac{1}{2}1;-\frac{1}{2}-\frac{1}{2}-1\biggl)\nonumber\\
  &~~~~~~~~~~~~~~~~~~~\times~~\prescript{}{T=1}{\biggl\langle} p(1)p(2)\biggl|\frac{1}{2}({\bf T}^2-\widehat{t}^2(1)-\widehat{t}^2(2))\biggl|p(1)p(2)\biggl\rangle_{T=1}\nonumber\\
  &=\frac{1}{4}~.
  \end{align}
In the above equation "$C$" denotes the Clebsch–Gordan (CG) coefficients \cite{Rose}.
The other matrix elements can be similarly calculated. Thus, the proton-proton, neutron-neutron, and the parts of proton-neutron matrix elements with $T=1$ can be written as

\begin{eqnarray}
 \widehat{v}_{ijkl,J}^{(pp)}=\widehat{v}_{ijkl,JT=1}^{(0)}-\frac{1}{2}\widehat{v}_{ijkl,JT=1}^{(1)}+\frac{1}{6}\widehat{v}_{ijkl,JT=1}^{(2)}
  ~~~~~\textrm{or}~~~\widehat{v}^{(pp)}=\widehat{v}^{(0)}-\frac{1}{2}\widehat{v}^{(1)}+\frac{1}{6}\widehat{v}^{(2)}~,
  \label{vpp}
\end{eqnarray}
where 
\begin{eqnarray}
 \widehat{v}_{ijkl,J}^{(pp)}=\sum_{k}\biggl\langle ij;J\biggl|v^{(k)}\biggl|kl;J\biggl\rangle ~\prescript{}{T=1}{\biggl\langle} p(1)p(2)\biggl|I^{(k)}\biggl|p(1)p(2)\biggl\rangle_{T=1}~.\nonumber
   \label{vppdef}
\end{eqnarray}
Similarly,
\begin{eqnarray}
  \widehat{v}^{(nn)}=\widehat{v}^{(0)}+\frac{1}{2}\widehat{v}^{(1)}+\frac{1}{6}\widehat{v}^{(2)}~,
  \label{vnn}
\end{eqnarray}
and
\begin{eqnarray}
  \widehat{v}^{(pn)}=\widehat{v}^{(0)}-\frac{1}{3}\widehat{v}^{(2)}~.
  \label{vpn}
\end{eqnarray}
Adding and subtracting (\ref{vpp}), (\ref{vnn}) and  (\ref{vpn}), we obtain
\begin{eqnarray}
 \widehat{v}^{(pp)}+\widehat{v}^{(nn)}+\widehat{v}^{(pn)}=3\widehat{v}^{(0)}~~~~~
  \textrm{or}~~~\widehat{v}^{(0)} =\frac{1}{3}\left(\widehat{v}^{(pp)}+\widehat{v}^{(nn)}+\widehat{v}^{(pn)}\right)~,
\end{eqnarray}
\begin{eqnarray}
  \widehat{v}^{(1)} =\widehat{v}^{(nn)}-\widehat{v}^{(pp)}~,
  \label{H1}
\end{eqnarray}
\begin{eqnarray}
  \widehat{v}^{(2)} =\widehat{v}^{(pp)}+\widehat{v}^{(nn)}-2\widehat{v}^{(pn)}~.
  \label{H2}
\end{eqnarray}
It is now evident from Eqs. (\ref{H1}) and (\ref{H2}) that $\widehat{v}^{(1)}$ and $\widehat{v}^{(2)}$ are responsible for CSB and CIB in the nucleon-nucleon interaction, respectively.

The  Hamiltonian given by Eq. (\ref{HTOT1}) can be expressed in terms of charge-independent and charge-violating parts as \cite{ORMAND19891}
\begin{eqnarray}
\widehat{H}_{\textrm{tot}} =\widehat{H}_{CI}+\widehat{H}_{CV}^\prime ~,
  \label{Hamil}
\end{eqnarray}
and the binding energy is given by

\begin{eqnarray}
   \textrm{BE}(\alpha T T_z)=\langle \alpha T T_z|(\widehat{H}_{CI}+\widehat{H}_{CV}^{\prime})|\alpha T T_z\rangle~.
  \label{BE11}
\end{eqnarray}
If the structure of $\widehat{H}_{CV}^{\prime}$, involves only two-body interactions, we can write it as a tensor upto rank two as
\begin{eqnarray}
\widehat{H}_{CV}^{\prime} =\sum_{k=0}^2 \widehat{H}_{CV}^{(k)}~,
\end{eqnarray}
where $k=0,1$ and 2 correspond to
\begin{eqnarray}
\widehat{H}_{CV}^{(0)} =\frac{1}{3}\left(\widehat{v}^{(pp)}+\widehat{v}^{(nn)}+\widehat{v}^{(pn)}\right)~,
\end{eqnarray}
\begin{eqnarray}
  \widehat{H}_{CV}^{(1)} =\widehat{v}^{(nn)}-\widehat{v}^{(pp)}~,
\end{eqnarray}
\begin{eqnarray}
  \widehat{H}_{CV}^{(2)} =\widehat{v}^{(pp)}+\widehat{v}^{(nn)}-2\widehat{v}^{(pn)}~.
\end{eqnarray}

In the following, we shall now obtain IMME of Wigner by considering the contribution of $\widehat{H}_{CV}^\prime$ in the first order perturbation
theory. This contribution gives the energy splitting of the isobaric mass multiplets, i.e.,
\begin{eqnarray}
  \Delta \textrm{BE}(\alpha T T_z)=\biggl\langle \alpha T T_z\biggl|\sum_{k=0}^2\widehat{H}_{CV}^{k} \biggl|\alpha T T_z\biggl\rangle
 =\langle \alpha T T_z| (\widehat{H}_{CV}^{(0)}+\widehat{H}_{CV}^{(1)}+\widehat{H}_{CV}^{(2)})|\alpha T T_z\rangle~.
\end{eqnarray}
The explicit $T_z$ dependence of the energy splitting of the multiplet can  be factored out using the Wigner-Eckart theorem
\cite{Rose} as 
\begin{align}
  \Delta \textrm{BE}(\alpha T T_z)
  &=C(T0T;T_z 0T_z)
  \langle \alpha T ||\widehat{H}_{CV}^{(0)} ||\alpha T\rangle  
  +C(T1T;T_z 0T_z)
  \langle \alpha T ||\widehat{H}_{CV}^{(1)} ||\alpha T\rangle  \nonumber\\\nonumber\\
  &~~~~~~~~~~~~~~~~~~~~~~~~~~~~+C(T2T;T_z 0T_z)
  \langle \alpha T ||\widehat{H}_{CV}^{(2)} ||\alpha T\rangle ~,\nonumber\\\nonumber\\
  &=\langle \alpha T ||\widehat{H}_{CV}^{(0)} ||\alpha T\rangle  
  +\frac{T_z}{[T(T+1)]^{1/2}}
  \langle \alpha T ||\widehat{H}_{CV}^{(1)} ||\alpha T\rangle  \nonumber\\\nonumber\\
  &~~~~~~~~~~~~~~~~~~~~~~~+\frac{3T_z^2-T(T+1)}{[T(T+1)(2T-1)(2T+3)]^{1/2}}
  \langle \alpha T ||\widehat{H}_{CV}^{(2)} ||\alpha T\rangle~,
  \label{DBE}
\end{align}
where the algebraic expressions of the Clebsch-Gordan coefficients have been used \cite{Edmonds}. The double-bars in the above
equation denote the isospin reduced matrix elements that depend on the total isospin and any other quantum number necessary to define the states
uniquely. The expression on the right side of Eq. (\ref{DBE}) acquires the following quadratic form by rearranging the terms:
\begin{eqnarray}
  \Delta \textrm{BE}(\alpha, T, T_z)~=~a~+~b~T_z~+~c~T_z^2~.
  \label{deltaBE}
\end{eqnarray}
The coefficient "$a$" depends on the isoscalar part of the interaction with a small contribution from isotensor component, coefficient
"$b$" is related to the isovector and "$c$" on isotensor parts of the nucleon-nucleon interaction. These parameters are fitted to the
empirical data and include the effects of the Coulomb term and also the ISB term of the nucleon-nucleon interaction. In the
derivation of Eq.~(\ref{DBE}), it is assumed that ISB term of the nucleon-nucleon interaction has the same structure as that
of the Coulomb interaction and it is an interesting problem to derive the isospin dependence of the nucleon-nucleon interaction starting
from QCD inspired microscopic theories \cite{Sagawa:2023itk}.


It is possible to have a rough estimate of the three parameters in Eq.~(\ref{IMME1}) by assuming a uniform charged spherical distribution
for the nucleus, and the Coulomb energy is then simply given by \cite{Bentley2007} 
\begin{eqnarray}
  E_C=\frac{3e^2Z(Z-1)}{5R_C}
  =\frac{3e^2}{5r_0A^{\frac{1}{3}}}\left[\frac{A}{4}(A-2)+(1-A)T_z+T_z^2\right]
  =a+bT_z+cT_z^2~,
  \label{IMME1}
\end{eqnarray}
with
\begin{eqnarray}
  a=\frac{3e^2A(A-2)}{20r_0A^{\frac{1}{3}}},~~b=-\frac{3e^2(A-1)}{5r_0A^{\frac{1}{3}}},~~c=\frac{3e^2}{5r_0A^{\frac{1}{3}}}~.
  \label{abc1}
\end{eqnarray}

Recent advancements in nuclear physics have led to a reevaluation of the IMME, particularly, in light of newly available data regarding nuclei with $T \ge 3/2$ \cite{Bentley2007}. These experimental endeavours, while promising, present formidable challenges due to the necessity for exceptionally precise mass measurements of analog nuclei that are notably deficient of neutrons. Despite decades of effort, achieving the requisite level of precision has proven to be a gradual process.

\begin{figure}[htp]
 \centerline{\includegraphics[trim=0cm 0cm 0cm
0cm,width=0.5\textwidth,clip]{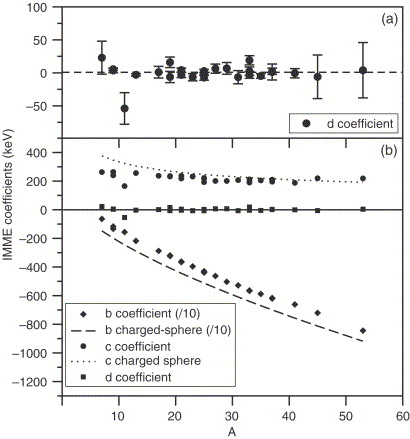}} \caption{(Color
   online) Coefficients of the IMME of the form $\Delta\textrm{BE}(T,T_z)=a+bT_z+cT_z^2+dT_z^3$ obtained from fits to experimental data for $T=\frac{3}{2}$ quadruplets 
   . (a) The experimentally determined $d$ coefficients, expected to be zero. (b) The experimentally determined $b$, $c$ and $d$ coefficients along with predictions of the simple charged sphere model. (Adopted from Ref. \cite{Bentley2007}).
  }
\label{Fig2}
\end{figure}

Fig.~\ref{Fig2} displays results of some recent studies where researchers have meticulously fitted known data to the IMME, employing
up to cubic term in $T_z$ to explore the potential necessity for higher-order correction terms in order to test the predictive power of
the model \cite{Bentley2007}. The analysis reveals interesting insight that with the exception of the isobar $A=9$, the data agrees
remarkably well with the quadratic IMME expression originally proposed 
by Wigner \cite{Ewigner1957}. Moreover, coefficients derived for a uniform spherical charge distribution, as depicted in Fig.~(\ref{Fig2}), exhibit a notable agreement with the fitted values.
 
The IMME has been derived by considering tensorial representation of the Coulomb interaction. However, since the parameters
of the mass formula are fitted to the data, it also includes the contributions from the ISB  terms of the
nuclear potential. It is, therefore, implicitly assumed that ISB terms of the nuclear potential have also  the same tensorial
form as that of the Coulomb interaction.
In a seminal investigation \cite{Dong2018}, the author tried to deduce the ISB contribution to the effective two-body 
interaction using a microscopic approach by  employing the bare Argonne AV18 potential. The medium effects were incorporated
by solving the Bethe-Goldstone equation within the 
Brueckner approach. In the following, we discuss some important results from this work \cite{Dong2018}.
In nuclear matter approximation, energy density with the inclusion of CSB and CIB terms is given by
\begin{eqnarray}
E(\rho ,\beta ) =E(\rho ,0)+S_{0}^{\text{(CIB)}}(\rho )+S_{1}^{\text{(CSB)}%
}(\rho )\beta+\left[ S_{2}(\rho )+S_{2}^{\text{(CIB)}}(\rho )\right] \beta ^{2}+%
\mathcal{O(}\beta ^{3})~, \label{EOS}
\end{eqnarray}
It is evident from above that CSB term of the nucleon-nucleon interaction gives the first order symmetry energy coefficient, whereas the
CIB contributes to the even orders only.
In the above equation, the density of nuclear matter is $\rho =\rho _{n}+\rho _{p}$ and the parameter $\beta =(\rho _{n}-\rho _{p})/\rho$
denotes the isospin asymmetry.
The derivation of Eq.  (\ref{deltaBE})  is based on Wigner's assumption of $|\alpha T T_z\rangle$  as the eigenstate of the charge-independent Hamiltonian  $\widehat{H}_0$, 
where $\alpha$ encompasses all additional quantum numbers to uniquely specify this state. All charge-violating two-body interactions, including the Coulomb interaction 
$\widehat{H}_{\text{C}}$ among protons and $\widehat{H}_{\text{CSB+CIB}}$ of CSB and CIB interactions, are treated as first-order perturbations. The total binding energy is expressed as
\begin{equation}
-\text{BE}(\alpha TT_{z})=\langle \alpha TT_{z}|(\widehat{H}_{0}+\widehat{H}_{\text{C}}+\widehat{H}_{\text{CSB+CIB}})|\alpha TT_{z}\rangle~,
\end{equation}
where  the first-order perturbation terms  $\widehat{H}_{\text{C}}$ and $\widehat{H}_{\text{CSB+CIB}}$  are expressed as tensor operators of up to
rank two. The matrix elements of tensor operators can be written in terms of  reduced matrix 
elements using the Wigner-Eckart theorem, with the coefficients involving solely depending on $T$ and $T_z$ and also
on radial and spin components. However, in a nuclear medium, $\widehat{H}_{\text{CSB+CIB}}$ becomes a density-dependent effective 
interaction, and can no longer be represented as an irreducible tensor \cite{Dong2018}. The perturbation
energy term corresponding to CSB and CIB components, i.e., $\langle \alpha
TT_{z}|\widehat{H}_{\textrm{CSB+CIB}}|\alpha TT_{z}\rangle$, lacks analytic forms unlike the Coulomb interaction. In the presence of effective CSB and CIB interactions, the perturbed energy can be represented as \cite{Dong2018}
\begin{equation}
\langle \alpha TT_{z}|H_{\text{CSB+CIB}}|\alpha TT_{z}\rangle =a_{\text{sym,1}%
}^{\text{(CSB)}}(A,T_{z})IA+a_{\text{sym,2}}^{\text{(CIB)}}(A,T_{z})I^{2}A~,
\end{equation}
where the coefficients "$a$" contain the zeroth order symmetry contribution
and $I = (N - Z)/A = 2T_z/A$ . The generalized IMME (GIMME), incorporating the perturbation correction from CIB and CSB terms, is then expressed as 
\begin{eqnarray}
\text{ME}(A,T,T_{z})=a+\left( b_c+\Delta _{\text{nH}}+2a_{\text{sym,1}}^{\text{(CSB)}}(A,T_{z})\right) T_{z} +\left(c_{c}+\frac{4}{A}a_{\text{sym,2}}^{\text{(CIB)}}(A,T_{z})\right)T_{z}^{2}~,
 \label{HH}
\end{eqnarray}
where $\Delta _{\text{nH}}=0.782$ MeV is the mass difference between a neutron and a hydrogen nucleus. The above derivation disentangles 
 the effects of effective charge-violating nuclear interactions from those occurring due to the Coulomb potential. The terms
 like $b_c$ and $c_c$ in  Eq.~(\ref{HH}) are solely induced by the Coulomb interaction and are independent of the total isospin projection $T_z$. However, novel components such as $a^{\text{(CSB)}}_{\text{sym,1}}(A,T_z)$ and $a^{\text{(CIB)}}_{\text{sym,2}}(A,T_z)$, originating
 from CSB and CIB interactions in the nuclear medium involve $T_{z}$ dependence. This dependence of the coefficients signifies that
 the microscopically derived IMME cannot be expressed as a simple polynomial in $T_z$. 

\begin{figure}[H]
\begin{center}
\includegraphics[width=1.0\textwidth]{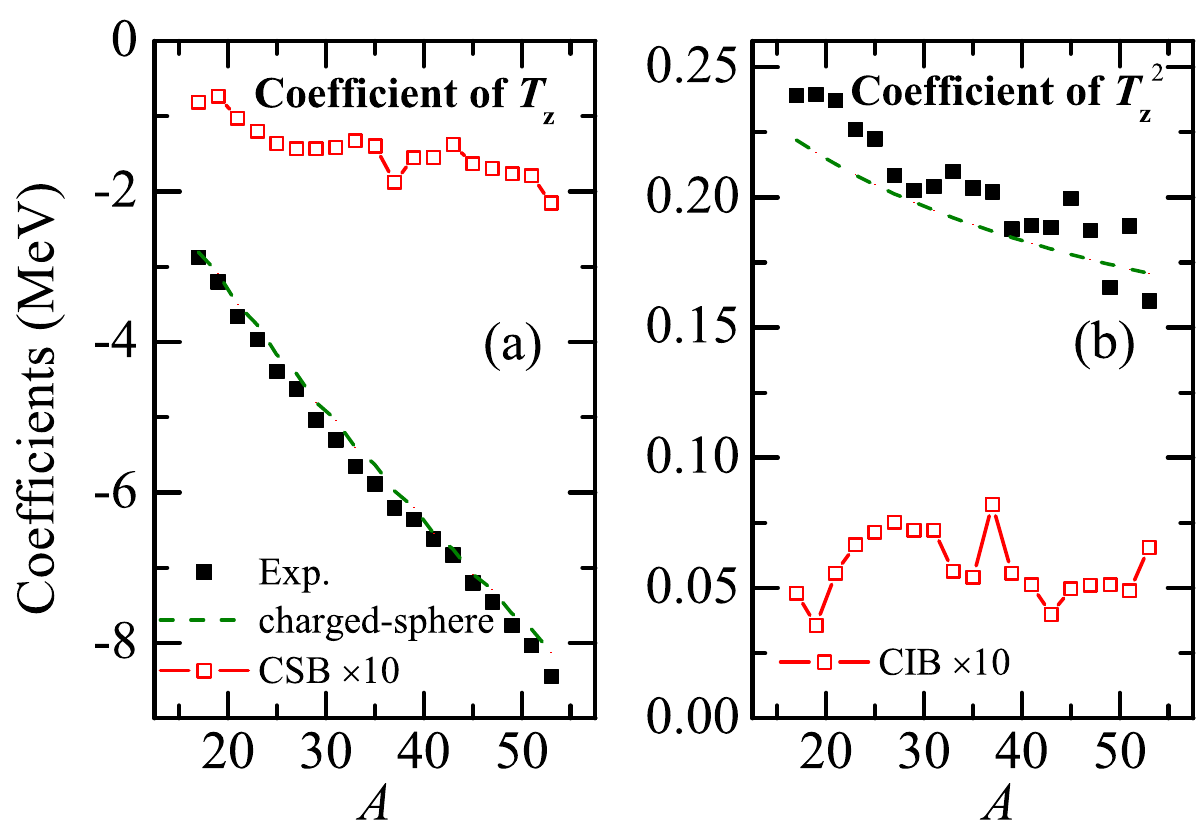}
\caption{For
the $T=3/2$ quartets, coefficients of (a) $T_z$ and (b)
  $T^2_z$ 
  extracted from the experimental data~\cite{MacCormick2014}, along with the calculated coefficients of the Coulomb
contribution (plus $\Delta _{\text{np}}$) from a simple
non-uniformly charged sphere 
(dashed curves) are shown for
comparison.  (The figure is adopted from Ref.~\cite{Dong2018}).
}\label{fig2}
\end{center}
\end{figure}

 The coefficients associated with the $T_z$ and $T_z^2$ terms of the IMME, deduced from the measured masses of  $T=3/2$ isobaric quartets ~\cite{MacCormick2014}, 
 are depicted in  Fig. \ref{fig2}, compared with those obtained from the model of a non-uniformly charged sphere ~\cite{Danielewicz2003}.
 Furthermore, Fig. \ref{fig2} illustrates the contributions stemming 
 from CSB and CIB effects in Eq. (\ref{HH}) for nuclei with  $T_z= T$ as illustrative examples.
 The analysis reveals that the contribution attributed to the CSB effect to the coefficient of the $T_z$ term
 rises  approximately from $-80$ keV to $-220$ keV as the mass number $A$ increases from 
17 to 53 \cite{Dong2018}. This observation aligns well with estimations for $T=1$ multiplets documented in Table 5.4 of Ref. \cite{Auerbach1972}. Broadly speaking, the CSB effect leads to a diminution of 
the $T_z$ term coefficient by  $2.0\%-3.1\%$, while the CIB effect augments the coefficient of the $T_z^2$ term by $1.6\%-4.4\%$.
It is noteworthy that although the energy splitting within the isobaric multiplet primarily arise from the Coulomb interaction, the corrections to the IMME clearly have CSB and CIB origins.
 \subsection{Nolen-Schiffer anomaly}
 The energy splitting in an isobaric mass multiplet for a given angular-momentum and parity should be equal to the difference in the
 Coulomb energies and be described by the Wigner's IMME. However, it has been known for more than fifty year that even
 after including a string of correction terms, which includes the electromagnetic
 spin-orbit interactions, an overall shift from a few to ten percent in the Coulomb displacement energies (CDE) still remains unresolved.
 This difference is known as Nolen-Schiffer anamoly \cite{Nolen1969, Auerbach1983}.
 Using IMME, CDE for adjacent members of a multiplet is given by [Ref. \cite{Dong2018}]
\begin{eqnarray}
  \textrm{CDE}(A,T,T_z)=-b-c(2T_z+1)+\Delta_{\textrm{nH}}~,
\end{eqnarray}
where $T_z$ is taken for the isobar with larger proton number. With the GIMME, the above expression is modified as
\begin{eqnarray}
  \textrm{CDE}(A,T,T_z)=-b_c-c_c(2T_z+1)+\Delta_{\textrm{NSA}}~,
\end{eqnarray}
where $\Delta_{NSA}$ is arising from CSB and CIB and is given by [Ref. \cite{Dong2018}]
\begin{align}
  \Delta_{\textrm{NSA}}=&-2a_{sym,1}^{\textrm{(CSB)}}(A,T_{z>})-\frac{4(2T_z+1)}{A}a_{sym,2}^{\textrm{(CIB)}}(A,T_{z>})+2T_z\bigg[a_{sym,1}^{\textrm{(CSB)}}(A,T_{z})-a_{sym,1}^{\textrm{(CSB)}}(A,T_{z>})\bigg]\nonumber\\
 & ~~~~~~~~~~~~~~~~~~~~~~~~~~~~~~~~+\frac{4T_z^2}{A}\bigg[a_{sym,2}^{\textrm{(CIB)}}(A,T)-a_{sym,2}^{\textrm{(CIB)}}(A,T_{z>})\bigg]\nonumber\\
  \approx&-2a_{sym,1}^{\textrm{(CSB)}}(A,T_{z>})-\frac{4(2T_z+1)}{A}a_{sym,2}^{\textrm{(CIB)}}(A,T_{z>})~,
\end{align}
with $T_{z>}=T_z+1$. It is evident from above that $\Delta_{NSA}$ has contribution from CSB and CIB apart from the usual Coulomb perturbation \cite{Dong2018}. $\Delta_{NSA}$ accounts for about $2\%$ to $3\%$ of the CDE for isobaric quartets.

\section{ Nuclear energy density functional approach and isospin symmetry breaking }
\label{sec:iso_dft}
In this section, we shall employ the nuclear energy density functional (NEDF) approach to investigate the isospin mixing
for the nuclear many-body system. This section is subdivided into five subsections. In subsection (\ref{sec:iso_dft_1}), a general outline
of the NEDF approach is presented from historical perspective. In subsection (\ref{sec:iso_dft_2}), generalized NEDF model is
presented that preserves isospin-symmetry at the DFT level and includes densities in all isospin channels. An approximate
isospin projection method is discussed in subsection (\ref{sec:iso_dft_3}). In subsection (\ref{sec:iso_dft_4}), ISB terms are introduced
in NEDF method and finally in subsection (\ref{sec:iso_dft_5}) numerical results of a few isobaric chains of isotopes
are presented and discussed.

\subsection{Timeline of the DFT in nuclear physics}
\label{sec:iso_dft_1}
DFT has become an indispensable and a valuable framework to investigate the properties
of many-body quantum systems \cite{parr1989}. It is now widely used in physics, chemistry and material science to
simulate many-body correlations. The DFT approach is based on the two theorems of Hohenberg-Kohn \cite{Hohenberg1964}. The first of these theorems
states that every system has a unique local density that describes the ground-state energy of a system, and the second theorem
provides a procedure on how to obtain this density. However, the major impediment in the application of the
DFT approach is how to construct the energy density functional, in the first place, in terms of the local density that will lead to the exact
ground-state energy of the Schrodinger wave function approach. Hohenberg-Kohn approach only establishes that there is a local density for
each system that gives the same ground-state energy as that of the Schrodinger approach. The ground-state energy of a system in an external
potential, ${\widehat{v}_{ext}}$, is given by \cite{GCola2020} 
\begin{align}
  E(\rho) &= \langle  \Psi |{ \widehat H} | \Psi \rangle ~,\nonumber\\
         &=  F(\rho) + \int d^3r ~ {\widehat{v}_{ext}} ~ \rho({\bf r})~.
\end{align}
In the first line of the above equation, the energy is expressed in terms of the standard
Schrodinger wave function. The second line is based on the Hohenberg-Kohn theorem, which
expresses the ground-state energy completely in terms of the spatial density. $\widehat {v}_{ext}$ in the
above equation is the external potential and in the case of atomic and molecular systems, it is the
Coulomb interaction.
The quantity $F(\rho)$ is the universal energy density functional for a given
species of particles, and the problem with the Hohenberg-Kohn theory is that it doesn't provide
any procedure on how to construct it. A commonly adopted scheme proposed by Kohn and Sham \cite{Kohn1965}
is to express the density in terms of the single-particle orbits, $\phi_i(\vec r)$ as
\begin{equation}
\rho({\bf r}) = \sum_{j} | \phi_j({\bf r})|^2~.  
\end{equation}
The kinetic energy in terms of single-particle orbits is given by
\begin{equation}
T   =   \sum_{j}\int d^3r \phi_j^*({\bf r})\biggl(-\frac{\hbar^2}{2m}\nabla^2 \biggl)\phi_j({\bf r})~.
\label{KE1}
\end{equation}
The functional $F(\rho)$ corresponds to the Hamiltonian of the system with the kinetic energy contribution, evaluated in Eq.~(\ref{KE1}). For atomic systems, the direct component of Coulomb energy is
approximated by the classical electrostatic energy  and only the unknown component of the energy functional
is the exchange term.  

In nuclear physics, DFT has its roots in the 1970s and 80s pursuit to develop an effective interaction that will elucidate the properties of
nuclei all across the periodic table \cite{Bender2003}. In this class of
nucleon-nucleon forces, the interaction potential containing several terms is completely parameterized. The most successful of these effective
interactions is the Skyrme force introduced in late fifties \cite{Skyrme1956, SKYRME1958615, SKYRME195835}. It was shown that this short-range expansion of the two-body potential leads to a simple structure of the
Hamiltonian density in terms of the nuclear and kinetic densities in the Hartree-Fock (HF) approach. For the three-body force, Skyrme also
introduced a zero-range force and it was shown that it corresponds to a linear density-dependent two-body force in HF approximation for
time even systems. The most important advantage of the Skyrme potential is that HF equations can be derived directly without explicitly calculating
two- and three-body matrix elements of the potential. The other advantage of the Skyrme interaction is that
analytic expressions of the parameters are obtained in the
nuclear matter approximation. In the pioneering
work \cite{Vautherin1972}, HF calculations were performed
for doubly-closed shell  systems and it was demonstrated that binding energies and radii are reproduced quite accurately with two
different parameter sets obtained in this work. The original parameter sets of the Skyrme interaction, referred to as SI and SII, were obtained
by fitting the binding energy and density of nuclear matter, and also the binding energy and radii of $^{16}$O and $^{208}$Pb closed shell nuclei.

In a significant development, new parameter sets of the Skyrme interaction were obtained by fitting the coupling constants to a larger body of
data on closed and semi-closed nuclei \cite{Beiner1975}. In this work, binding energies and charge radii of
$^{16}$O, $^{40}$Ca, $^{48}$Ca, $^{56}$Ni, $^{90}$Zr, $^{140}$Ce and $^{208}$Pb were  used in the fitting procedure and several new sets
of Skyrme parameters were obtained. These parameter sets called as SIII, SIV, SV and SVI correspond to varying density dependencies of the interaction
potential. In this exhaustive work, around 200 nuclei were studied in the spherical HF+BCS approximation and it was shown that binding energies
and radii could be reasonably reproduced with all four new parameter sets. However, it was shown that calculated single-particle spectra deviate in the
four sets as stronger density dependence leads to a higher level density \cite{Beiner1975}.

The above sets of parameters from SI to SVI, referred to as the first generation, are rooted in the nuclear Hamiltonian with linear density
dependence. It was realised very early that although these sets describe the bulk properties quite well, but completely overestimate  the
energy of the isoscalar giant resonance \cite{Bender2003}. This mode related to the
incompressibility modulus, $K_{nm}$
\cite{TREINER1981253} and for SKIII, its value is more than 300 MeV and the empirical estimate from the isoscalar mode is about
220 MeV. It has been shown that in order to reproduce the empirical value, the density dependence of the Skyrme interaction should be between
$1/6$ and $1/3$ and a new Skyrme interaction set SkM was developed to reproduce the isoscalar monopole resonance
\cite{KRIVINE1980155, BARTEL198279}. The introduction of the fractional density dependence is a
complete departure from the linear dependence in the first generation of Skyrme force as this dependence cannot be derived from the
Hamiltonian of the system. However, it can be justified from the Hohnberg-Kohn DFT formulation as the energy density functional of a quantum
many-body system can be expressed in terms of any power (integer or fractional) of the local density. The Skyrme energy density in nuclear physics
,therefore, falls within the realm of DFT formulation of quantum mechanics as soon as the fractional power of the density is introduced.

In most of the DFT formulations in nuclear physics, beginning from 1990s, the energy density functional is constructed disregarding
the Hamiltonian origin of the new terms introduced in the functional. This new path in nuclear DFT was laid in the seminal
work \cite{REINHARD1995467} where
the spin-orbit interaction of the original Skyrme Hamiltonian was altered with the objective that anomaly found in the description
charge radii of the Pb-isotopes could be resolved. It was noted that Skyrme HF+BCS calculations with the standard parameter set fails
to reproduce the kink found in the charge radii of Pb-isotopes at N=126 \cite{TAJIMA1993434}. However, the HF+BCS
analysis performed with the relativistic density function, referred to as RMF approach \cite{RING1996193}, was capable of
reproducing the kink observed in the data. In the work \cite{REINHARD1995467}, effective non-relativistic energy density was
derived from the RMF approach and it was identified that spin-orbit term contained in this functional was different from that of spin-orbit interaction employed in the original Skyrme density functional. An auxiliary term in the spin-orbit part of the density functional was introduced and by refitting the
parameters, the kink  observed in the charge radii of Pb-isotopes was replicated in the Skyrme approach \cite{REINHARD1995467}.  

There is now a plethora of Skyrme energy density functional parameter sets and are obtained by a fitting coupling constants to ground-state properties
of magic and semi-magic nuclei, energies of giant resonances, fission barrier heights and the properties of symmetric and asymmetric nuclear matter.
In a comprehensive investigation \cite{Dutra2012}, 240 Skyrme functionals were scrutinized against 11 benchmark
empirical properties of nuclear matter ranging from low density dilute Fermi gas to a high density neutron matter. It was shown that out of
240 sets, only 16 were able to satisfy all the proposed constraints \cite{Dutra2012}. 

In recent years, modern optimization techniques have been developed to fit the Skyrme density functional coupling constants to a larger set of data, including those of deformed nuclei \cite{Kortelainen2010}. In this work, masses, charge radii
and odd-even mass differences of 72 nuclei ( 28 spherical and  44 deformed ) were selected in the fitting procedure. The derivative free optimization
algorithm POUNDerS \cite{Wild2015} was used and it was found to be much superior to the standard fitting algorithms.
It was shown that using the newly obtained parameter set, called as UNEDF0, returns root mean square (rms) deviations of 0.45 MeV and 1.2 Mev for
two-neutron separation energies and masses, respectively, of nuclei above A=80. However, the UNEDF0 functional was not designed to provide
a proper description of the single-particle energies as none of the properties related to shell structure were used in the
fitting procedure. In a subsequent work \cite{Kortelainen2012}, a new Skyrme functional, referred to as UNEDF1, was developed
by also including the excitation energies of fission isomers, observed in the actinides in the fitting procedure apart from the observables employed
in UNEDF0 stage.

In the third stage of the development of Skyrme functional following the UNEDF pathway, a new parameterization set so called UNEDF2
\cite{Kortelainen2014} was obtained by adding the tensor terms and including the data on
single-particle energies of doubly magic nuclei. Further, the data set used in UNEDF1 was extended in order to improve the pairing
properties in heavy mass region. However, it was realized  \cite{Kortelainen2014} that no significant
improvement in the overall description of the observed properties was achieved by employing more data in the fitting protocol. The
conclusion of the work \cite{Kortelainen2014} is reproduced here "However, after adding new data aiming to better constrain the nuclear functional, its quality has improved only marginally. These results suggest that the standard Skyrme energy density has reached its limits, and significant changes to the form of the functional are needed".

It is quite evident that Skyrme density functional needs to be enriched to achieve the quality of predictions of spectroscopic level. This statement
is also applicable to Gogny \cite{Decharg1980, Chappert2015} and relativistic \cite{RING1996193} density functionals as quality of agreement with data is similar to that achieved with
Skyrme functional. There are several pathways to extend the Skyrme density functional approach and quite obvious generalization is to include
the higher order derivative terms beyond the traditional quadratic form \cite{Raimondi2011}. There are also
attempts to augment the nuclear density functionals with long-range pion exchange contributions, obtained from the chiral
effective field theory \cite{Zurek2024}.

Further, it is known that existing NEDF have deficiencies that can be overcome in order to improve the
quality of predictions. For instance, the NEDF breaks the isospin symmetry as can be easily checked that energy functionals are expressed in terms
of isoscalar and only z-component of the isovector density, x- and y-components that exchange neutrons and protons are missing. The
density functional is therefore not scalar in the isospin space and breaks the isospin symmetry. This symmetry is quite important
for nuclei close to the $N \sim Z$ line and it is known that a phenomenological Wigner term needs to be added in the microscopic mass
models to correctly describe the extra binding energy measured for these systems \cite{Goriely2002, SAMYN2002142}. Further, in several studies, neutron-proton pairing term have been included in mean-field studies
\cite{Goodman1979, SATULA19971, FRAUENDORF1999509, TERASAKI19981} and the importance of the T=0 and T=1 pairing modes have been debated. In these
studies, the two pairing modes are mutually exclusive, however, in some exactly  solvable models, it turns out
that both modes can coexist \cite{Sheikh2000, Romero2019}.

Nevertheless, in all the above investigations, neutron-proton coupling has been considered only in the pairing channel and is
disregarded in the
particle-hole channel which clearly is an inconsistent approach that breaks the isospin symmetry.
In an attempt to develop a generalized DFT approach that obeys isospin symmetry in the particle-hole channel, the
nuclear energy density functional has been constructed \cite{Per04} that contains terms
up to quadratic order in terms of
local isoscalar and the three components of the isovector densities. The energy density functional has been constructed
by considering all possible terms that leave the functional invariant under spatial rotations, isospin rotations and space
inversion. In the following, we shall provide some details about this generalized approach and shall closely follow
the Refs. \cite{Per04, Sato2013, Sheikh2014}.

\subsection{Isospin invariant nuclear DFT}
\label{sec:iso_dft_2}
In the generalized isospin invariant approach, neutrons and protons are treated as two states of the same particle and
the ground-state wavefunction in the Kohn-Sham (Hartree-Fock) approximation for a system of "$A$" nucleons is given by 
\begin{equation}
 |\Psi\rangle=  \prod_{k=1}^{A}c_k^\dagger|0\rangle~,
\end{equation}
where the Kohn-Sham orbitals are related to the original fermion states through a unitary transformation
\begin{equation}
 c_k^\dagger=\int d^3{\bf r} \sum_{st}\phi_k^*({\bf r}st)a^\dagger_{{\bf r}st}~~~~~~~~(k\le A)~.
\end{equation}
The particle density matrix in the isospin representation is given by
\begin{equation}
 \widehat{\rho}({\bf r}st,{\bf r}^\prime s^\prime t^\prime )=\langle \Psi|a^\dagger_{{\bf r}^\prime s^\prime t^\prime}a_{{\bf r}st}|\Psi \rangle=\sum_{k=1}^{A}\phi_k({\bf r}^\prime s^\prime t^\prime )\phi_k^*({\bf r}st)~,
\end{equation}
and the trace of the above density gives the total particle number "$A$". The non-local scalar and vector densities
in the coordinate space can be written in terms of the above density matrix as
\begin{equation}
\rho_m({\bf r},{\bf r}^\prime)=\sum_{stt^\prime}\widehat{\rho}({\bf r}st,{\bf r}^\prime st^\prime)~\widehat{\tau}^m_{t^\prime t}~,
\end{equation}
\begin{equation}
{\bf s}_m({\bf r},{\bf r}^\prime)=\sum_{ss^\prime tt^\prime}~\widehat{\rho}({\bf r}st,{\bf r}^\prime s^\prime t^\prime)~\widehat{\sigma}_{s^\prime s}\widehat{\tau}^m_{t^\prime t}~,
\end{equation}
where $\widehat {\sigma}$ and $\widehat {\tau}$ are the Pauli matrices in the spin and isospin spaces, respectively. In the above
equation, $m=0$
corresponds to the isoscalar component, and $m=1,2$ and 3 are the three isovector components. The energy of the nuclear system in terms
of Skyrme density functional and the Coulomb terms is given by
\begin{equation}
  \overline{H}[ \widehat{\rho}]~=~\int d^3{\bf r}~\mathcal{H}({\bf r})=\int d^3{\bf r}~\mathcal{H}_{Sk}({\bf r})+
  E_{\textrm{Coul}}[\widehat{\rho}]~.
  \label{HKSHFP}
\end{equation}
The Skyrme energy density functional is given by
\begin{equation}
\mathcal{H}_{Sk}({\bf r})=\frac{\hbar^2}{2m}\tau_0({\bf r})+\chi_0({\bf r})+\chi_1({\bf r})~,
\end{equation}
where the first term is the kinetic energy density. The second and third terms are isoscalar and isovector parts of the
Skyrme interaction, respectively and when expressed as a functional of local densities and considering only terms that are
invariant under spatial and isospin rotations, these terms can be written as
\begin{align}
  \chi_0({\bf r})=&C_0^\rho \rho_0^2+C_0^{\Delta\rho} \rho_0\Delta\rho_0+C_0^\tau\rho_0\tau_0+C_0^{J0}{\bf J}_0^2+C_0^{J1}{\bf J}_0^2+C_0^{J2}{\bf J}_0^2+C_0^{\nabla J}\rho_0 {\bm \nabla}.{\bf J}_0+C_0^s{\bf s}_0^2\nonumber\\
  &+C_0^{\Delta s}{\bf s}_0.\Delta {\bf s}_0+C_0^T {\bf s}_0.{\bf T}_0+C_0^j{\bf j}_0^2+C_0^{\nabla j}{\bf s}_0.(\bm \nabla\times {\bf j}_0)+C_0^{\nabla s}(\bm \nabla . {\bf s}_0)^2+C_0^F {\bf s}_0.{\bf F}_0\\
  \chi_1({\bf r})=&C_1^\rho {\bm\rho}^2+C_1^{\Delta\rho} {\bm \rho}\otimes \Delta{\bm \rho}+C_1^\tau{\bm \rho}\otimes{\bm\tau}+C_1^{J0}{\bf J}^2+C_1^{J1}{\bf J}^2+C_1^{J2}{\bf J}^2+C_1^{\nabla J}{\bm \rho}\otimes {\bm \nabla}.{\bf J}+C_1^s{\bf s}^2\nonumber\\
  &+C_1^{\Delta s}{\bf s}.\otimes\Delta {\bf s}+C_1^T {\bf s}.\otimes{\bf T}+C_1^j{\bf j}^2+C_1^{\nabla j}{\bf s}.\otimes(\bm \nabla\times {\bf j})+C_1^{\nabla s}(\bm \nabla . {\bf s})^2+C_0^F {\bf s}.\otimes{\bf F}~,
\end{align}
where details on various terms and densities can be found in the
Refs. \cite{Per04, Sato2013, Sheikh2014}.
The Kohn-Sham or Hartree-Fock potential is obtained through variation of the energy density functional of Eq.~(\ref{HKSHFP}), i.e.,
\begin{align}
  \widehat{h}({\bf r}^\prime s^\prime t^\prime,{\bf r}st)&=\frac{\delta\overline{H}[\widehat{\rho}]}{\delta\widehat{\rho}({\bf r}st,{\bf r}^\prime s^\prime t^\prime)}\nonumber\\
  &=-\frac{\hbar^2}{2m}\delta({\bf r}-{\bf r}^\prime){\bm \nabla}. {\bm \nabla}\delta_{s^\prime s}\delta_{t^\prime t}+\widehat{\Gamma}({\bf r}^\prime s^\prime t^\prime,{\bf r}st)+\widehat\Gamma_r({\bf r}^\prime s^\prime t^\prime,{\bf r}st)~,
\end{align}
or
\begin{equation}
  \widehat{h}({\bf r}^\prime s^\prime t^\prime,{\bf r}st)=\delta({\bf r}-{\bf r}^\prime)\widehat{h}({\bf r};s^\prime t^\prime ,st)~,
\end{equation}
with
\begin{equation}
  \widehat{h}({\bf r};s^\prime t^\prime ,st)=h_0({\bf r};s^\prime ,s)\delta_{t^\prime ,t}+{\bf h}({\bf r};s^\prime ,s)\otimes\widehat{\bm \tau}_{t^\prime ,t}~,
\end{equation}
where
\begin{align}
  \widehat{h}({\bf r};s^\prime ,s)=&-\frac{\hbar^2}{2m}{\bm\nabla}^2\delta_{s^\prime s}\delta_{k0}+U_k\delta_{s^\prime s}+{\bm\Sigma}_k.\widehat{\bm \sigma}_{s^\prime s}+\frac{1}{2\iota}[{\bf I}_k\delta_{s^\prime s}+({\bf B}_k.\widehat{\bm \sigma}_{s^\prime s})].{\bm \nabla}\nonumber\\
 &+ \frac{1}{2\iota}{\bm \nabla}.[{\bf I}_k\delta_{s^\prime s}+({\bf B}_k.\widehat{\bm \sigma}_{s^\prime s})]-{\bm \nabla}.[{\bf M}_k\delta_{s^\prime s}+{\bf C}_k.\widehat{\bm \sigma}_{s^\prime s}]{\bm \nabla}-{\bm \nabla}.{\bf D}_k\widehat{\bm \sigma}_{s^\prime s}.{\bm \nabla}~,
\end{align}
and 
\begin{equation}
  ({\bf B}.\widehat{\bm\sigma})_a=\sum_b{\bf B}_{ab}\widehat{\bm\sigma}^b~,
\end{equation}
\begin{align}
  U_k({\bf r})&=2C_t^\rho \rho_k+2C_t^{\Delta\rho}\Delta\rho_k+C_t^\tau \tau_k+C_t^{\nabla J}{\bm \nabla}.{\bf J}_k~,\\
  {\bm\Sigma}_k{(\bf r)}&=2C_t^s{\bf s}_k+2(C_t^{\Delta s}-C_t^{\nabla s})\Delta{\bf s}_k-2C_t^{\nabla s}{\bm \nabla}\times ({\bm \nabla}\times {\bf s}_k)\nonumber\\
  &~~~~~~~~~~~~~~~~~~~~~~~~~~~~~~~+C_t^T {\bf T}_k +C_t^F {\bf F}_k+C_t^{\nabla j} {\bm \nabla}\times {\bf j}_k~,
\end{align}
\begin{align}
  {\bf I}_k({\bf r})&=2C_t^j {\bf j}_k+C_t^{\nabla j}{\bm\nabla}\times{\bf s}_k~,\\
  {\bf B}_k({\bf r})&=2C_t^{J0} {\bf J}_k\delta-2C_t^{J1}{\bm \epsilon}. {\bf J}_k+2C_t^{J2} {\bf J}_k+C_t^{\nabla J}{\bm \epsilon}.{\bm\nabla}\rho_k~,\\
  {\bf M}_k({\bf r})&=C_t^\tau \rho_k~,\\
  {\bf C}_k({\bf r})&=C_t^T {\bf s}_k~,\\
  {\bf D}_k({\bf r})&=C_t^F {\bf s}_k~.
\end{align}
Kohn-Sham orbitals are obtained through the solution of the following mean-field equations
\begin{equation}
  \int d^3{\bf r}\sum_{st}\widehat{h}({\bf r}^\prime s^\prime t^\prime ,{\bf r}st)~\phi_k^*({\bf r}st)=\varepsilon_k~\phi_k^*({\bf r}^\prime s^\prime t^\prime )~.
\end{equation}

It is known that the mean-field approximation breaks the isospin symmetry spontaneously and leads to spurious
isospin mixing in the nuclear states. This is especially true for systems with neutron excess and
necessitates that isospin projection of the HF states need to be performed.
As a matter of fact isospin projection was performed in early eighties \cite{CAURIER198011, CAURIER1982407} to
investigate the discrepancies reported in the HF approach to describe the Coulomb displacement
energies and differences in proton and neutron radii. In order to estimate the isospin mixing due to Coulomb interaction, the
Hamiltonian was diagonalized in the isospin projected states.

A procedure similar to the above diagonalization of the Hamiltonian has been adopted in the Skyrme DFT
approach \cite{Satula2009, Satula2010}. It has been shown that in comparison to the
angular-momentum and particle-number projection formalism in the DFT which are plagued with the singularity
problem \cite{Sheikh_2021}, the isospin projection
is free from this divergence. There are essentially two sources of divergences in nuclear DFT, one is due to
reason that nucleon vertices are not anti-symmetrized and secondly because of the application of the
generalized Wick's theorem \cite{Lacroix2009, Bender200979, Duguet200979}. These problems don't exist for nuclear density functionals derived from the Hamiltonian based
approaches, for instance, in the case of first generation of the Skyrme functionals. As already discussed earlier,
in all the modern nuclear DFT approaches, the functional terms are added with no linkages to the
nuclear Hamiltonian. It has been shown that nuclear DFT with integer power of density dependence, a regularization
approach is possible to rectify the divergence problem. However, for nuclear DFT with fractional density dependence,
regularization method is not possible and divergence problem persists \cite{Lacroix2009, Raja2020}.

The divergence occurs for the transition density as these are multiplied by the inverse of rotated norm kernel, which
has a pole structure for some rotational angles. In the case of isospin projection, the pole structure has
been investigated analytically and it has been shown that for the density dependence of the standard Skyrme
functional which has order less than three, the isospin projected energy density is free from spurious divergences
\cite{Satula2010}. 

Using the isospin projection approach, a chain of Ca-isotopes was studied with SLy4 interaction
and the isospin results
were compared before rediagonalization (BR) and after rediagonalization (AR) \cite{Satula2009}. It was noted that in the
absence of the Coulomb potential, the two results were
quite different with BR results showing no isospin mixing for the $N=Z$, but then increased with neutron excess.
This isospin mixing is spurious and occurs due to the spontaneous ISB by the mean-field
approximation. The results of AR depict no isospin mixing in the absence of Coulomb potential as expected in
an accurate treatment. In the presence of the Coulomb potential, the two results again differ with AR showing
quite large mixing for the $N=Z$ system and then dropping exponentially with neutron excess. The BR results
show smaller isospin mixing for the $N=Z$ system, but then remains non-zero even for the extreme neutron
excess system of $^{60}$Ca for which the AR shows zero mixing \cite{Satula2009}.

\subsection{Approximate isospin projection method}
\label{sec:iso_dft_3}
In the present work, we have followed an approximate isospin projection approach as outlined in Refs.
\cite{Sato2013, Sheikh2014}. In this scheme, the following auxiliary Kohn-Sham equation is solved $:$
\begin{align}
  \widehat{h}^\prime  =  \widehat{h}-{\bm\lambda}~.~\widehat{\bf t}
  =  \widehat{h}- \lambda_x~ \widehat{t}_x - \lambda_y~ \widehat{t}_y - \lambda_z~ \widehat{t}_z~.
  \label{AKSE}
\end{align}
In the above equation, three Lagrangian parameters $(\lambda_x, \lambda_y, \lambda_z)$ have been introduced that fix
the average values of the three components of the isospin, 
and is similar to the cranking model approach used in high-spin nuclear physics to generate the high angular momentum rotational
states. It is known that the cranking approach is a first order approximation to angular-momentum projection in the
Kamlah expansion \cite{ringschuck} of the projection operator. In Eq.~(\ref{AKSE}), the cranking term in introduced
in the isospace and, therefore, it will approximately project out the isospin quantum number. In the spherical
polar coordinate system,
$(\lambda_x, \lambda_y, \lambda_z) ~=~(\lambda \textrm{sin}\theta ~\textrm{cos}\phi,\lambda \textrm{sin}\theta ~\textrm{sin}\phi,\lambda \textrm{cos}\theta)$ and it has been shown that results are independent of the azimuthal angle ($\phi$) \cite{Sato2013}
and, therefore, all the calculations can be performed with two parameter of $\lambda$ and $\theta$. This prescription
ensures that $\langle \widehat{t_y} \rangle=0$ and corresponds to the invariance of the time-reversal symmetry in the
calculations.

The interpretation of the results becomes quite simple in the absence of the Coulomb interaction as
for this case, neutron-proton mixed DFT preserves the isospin and the states will have well defined
isospin quantum number. This symmetry invariance implies that a given "$T$" state should be degenerate for all the
isobaric analogue states which have different $T_z=(N-Z)/2$. On the two extremes of the isobaric multiplet,
the lowest states will have $T=T_z=|(N-Z)|/2$. Each single-particle Kohn-Sham energy state will
have neutron-proton and Kramers degeneracy. For the system having $T_z=(N-Z)/2$,
the lowest "$Z/2$" single-particle energies will be four-fold degenerate, and the remaining higher $(N-Z)/2$
single-particle energies will be two-fold degenerate. This system is solved using the standard DFT approach with
fixed neutron and proton numbers and the $\lambda$ parameters is chosen equal to
$\lambda=\lambda^{(T_z=T)}_{np} = (\lambda_n - \lambda_p)$ and $\theta$ is varied to obtain the solutions for analogue states of other isobars
\cite{Per04}. For $\theta=180^\circ$, protons and neutrons are interchanged and for values
of $\theta$ between $0^\circ$ and $180^\circ$, the states are mixed. $\theta=90^\circ$ has equal neutron and proton mixing
in the absence of ISB terms. Although neutron-proton mixing of the states varies with $\theta$, but energies remain constant.

The inclusion of the Coulomb interaction favours the states with larger $<T_z>$ as these states will have lower contribution
to the electrostatic energy. The single-particle energies now become "$\theta$" dependent and the level crossings occur. This
leads to ping-pong divergence problem in the HF iterative solution and it is difficult to find a converged solution.
To avoid this divergence, a shift in the $\lambda$ value is introduced \cite{Sato2013} in
the following manner :
\begin{eqnarray}
{\bm \lambda} = (\lambda \textrm{sin}\theta , 0, \lambda \textrm{cos}\theta) = (\lambda^\prime \textrm{sin}\theta^\prime , 0, \lambda^\prime \textrm{cos}\theta^\prime + \lambda_{\textrm{off}})~,
\end{eqnarray}
with
\begin{equation}
(\lambda_{\textrm{off}},\lambda^\prime) = \frac{1}{2} (\lambda_{np}^{T_z=T}+\lambda_{np}^{T_z=-T},\lambda_{np}^{T_z=T}-\lambda_{np}^{T_z=-T})~.
\end{equation}

\subsection{Isospin symmetry breaking in nuclear DFT}
\label{sec:iso_dft_4}
The ISB terms have been introduced in Skyrme density functional approach
\cite{BACZYK2018178} to shed light on the discrepancies obtained in the evaluation of
mirror displacement energies (MDEs) and triplet displacement energies (TDEs). It was observed that using
three different standard Skyrme functionals of SV$_{\textrm {T}}$, SkM$^{*}$ and SLy4, the MDEs for lighter mass nuclei were
underestimated by about $10\%$, and the observed staggering in the TDEs was completely missing in the calculated
values. It was shown that by introducing ISB terms of class II and III, the discrepancies in the
evaluation of MDEs and TDEs could be clarified \cite{BACZYK2018178}. These terms
were introduced in the three dimensional Skyrme DFT solver HFODD with neutron-proton mixing included at the HF level
\cite{SCHUNCK2017145}. In the present work, we have included these terms in the HFBTHO solver \cite{Sheikh2014} and in the following we shall
discuss this new development. The ISB terms introduced in the neutron-proton mixed HFBTHO program are :

\begin{eqnarray}
  \widehat{V}^{II}(i,j)=t_0^{II}\delta( {\bf r}_{i}-{\bf r}_{j})\biggl( 3\widehat{\tau}_{z}(i)\widehat{\tau}_{z}(j)-\widehat{{\bm \tau}}(i).\widehat{{\bm \tau}}(j)\biggl) ~,
\end{eqnarray}
\begin{eqnarray}
  \widehat{V}^{III}(i,j)=t_0^{III}\delta( {\bf r}_{i}-{\bf r}_{j})\biggl( \widehat{\tau}_{z}(i)+\widehat{\tau}_{z}(j)\biggl)~.
\end{eqnarray}
The corresponding contributions to NEDF are 
\begin{align}
  \langle \Phi| \widehat{V}^{II}|\Phi\rangle&=\frac{1}{2}\sum_{ij} \langle ij|t_0^{II}\delta({\bf r}_{i}-{\bf r}_{j})\biggl( 2\widehat{\tau}_{z}(i)\widehat{\tau}_{z}(j)-\widehat{\tau}_x(i)\widehat{\tau}_x(j)-\widehat{\tau}_y(i)\widehat{\tau}_y(j)\biggl)\nonumber\\
  &~~~~~~~~~~~~~~~~~~~~~~~~~~~~~~~~~~~~~~~~~~~~~~~~~~~~~~~~~~~~~~~~~~~~~~~~~~\biggl(1-\widehat{P}_{ij}^{M}\widehat{P}_{ij}^{\sigma}\widehat{P}_{ij}^{M}\biggl) |ij\rangle ~,\nonumber \\
  & =\frac{1}{2}t_0^{II}\sum_{ij} \langle ij|\delta({\bf r}_{i}-{\bf r}_{j}))\biggl( 2\widehat{\tau}_{3}(i)\widehat{\tau}_{3}(j)-\widehat{\tau}_1(i)\widehat{\tau}_1(j)-\widehat{\tau}_2(i)\widehat{\tau}_2(j)\biggl)\nonumber\\
  &~~~~~~~~~~~~~~~~~~~~~~~~~~~~~~~~~~~~~~~~~~~~~~~~~~~~~~~~~~~~~~~~~~~~~~~~~\frac{1}{2} \biggl(1-\boldsymbol{\sigma}(i).\boldsymbol{\sigma}(j) \biggl)|ij\rangle ~,\nonumber \\
  & =\frac{1}{4}t_0^{II}\int \biggl(2\rho_{z}^{2}(\boldsymbol{r})-\rho_{x}^{2}(\boldsymbol{r})-\rho_{y}^{2}(\boldsymbol{r})-2{\bf s}_{z}^{2}(\boldsymbol{r})+{\bf s}_{x}^{2}(\boldsymbol{r})+{\bf s}_{y}^{2}(\boldsymbol{r}) \biggl)d^3r~.
  \label{VIIEDF}
\end{align}
Defining
\begin{align}
  \rho_{0}(\boldsymbol{r})&=\rho_{nn}(\boldsymbol{r})+\rho_{pp}(\boldsymbol{r}) ~,\nonumber \\
  \rho_{x}(\boldsymbol{r})&=-(\rho_{np}(\boldsymbol{r})+\rho_{pn}(\boldsymbol{r})) ~,\nonumber \\
  \rho_{y}(\boldsymbol{r})&=-i(\rho_{np}(\boldsymbol{r})-\rho_{pn}(\boldsymbol{r})) ~,\nonumber \\
  \rho_{z}(\boldsymbol{r})&=\rho_{nn}(\boldsymbol{r})-\rho_{pp}(\boldsymbol{r}) ~,\nonumber \\
 \rho_{x}^{2}(\boldsymbol{r})&=\rho_{np}^{2}+\rho_{pn}^{2}+2\rho_{np}\rho_{np}~,\nonumber \\
 \rho_{y}^{2}(\boldsymbol{r})&=-(\rho_{np}^{2}+\rho_{pn}^{2}-2\rho_{np}\rho_{np})~,\nonumber \\
 \rho_{z}^{2}(\boldsymbol{r})&=\rho_{nn}^{2}+\rho_{pp}^{2}-2\rho_{nn}\rho_{pp} ~,\nonumber \\
 \rho_{x}^{2}(\boldsymbol{r})&+\rho_{2}^{2}(\boldsymbol{r})=4\rho_{np}\rho_{pn}~.
    \end{align}
 Using the above definition, Eq.~(\ref{VIIEDF}) become
   \begin{eqnarray}
  &  \langle \Phi| \widehat{V}^{II}|\Phi\rangle=\frac{t_0^{II}}{4}\int d^3r\biggl(2\rho_{nn}^{2}+2\rho_{pp}^{2}-4\rho_{nn}\rho_{pp}-4\rho_{np}\rho_{pn}-2{\bf s}_{nn}^{2}\nonumber\\
    & ~~~~~~~~~~~~~~~~~~~~~~~~~~~~~~~~~~~-2{\bf s}_{pp}^{2}+4{\bf s}_{nn}.{\bf s}_{pp}+4{\bf s}_{np}.{\bf s}_{pn}\biggl)~.
     \label{VIIEDF1}
   \end{eqnarray}
Similarly,    
\begin{align}   
   \langle \Phi| \widehat{V}^{III}|\Phi\rangle&=\sum_{ij}\langle ij|\delta({\bf r}_{i}-{\bf r}_{j})\biggl(\widehat{\tau}_3(i)+\widehat{\tau}_3(j)\biggl)\frac{1}{2}\biggl(1-\boldsymbol{\sigma}(i).\boldsymbol{\sigma}(j)\biggl)|ij\rangle\nonumber\\
   &=\frac{t_0^{III}}{2}\int d^3r\biggl(\rho_{z}\rho_{0}-{\bf s}_{0}.{\bf s}_{z}\biggl)\nonumber\\
   &=\frac{t_0^{III}}{2}\int d^3r\biggl(\rho_{nn}^2-\rho_{pp}^2-{\bf s}_{nn}^2+{\bf s}_{pp}^2\biggl)~,
\end{align}
   where $\rho$ and ${\bf s}$ are scalar and spin-vector densities, respectively. The expressions of particle and spin densities in the cylindrical
   coordinate system are provided in appendix (\ref{DENSITIES}). The cylindrical system is used in the HFBTHO solver as it employs the
   axial symmetry and the DFT problem reduces to two dimensions. The two coupling constants of the ISB terms have been separately
   fitted to the experimental values of MDEs and TDEs \cite{Baczyk2015}
   since their contributions have been shown to be of the order of few MeVfm$^3$ in comparison to the isoscalar coupling constant
   of the Skyrme functional, $t_0$, which is about 1000~MeVfm$^3$. It is noted from Eq.~(\ref{VIIEDF1}) that contribution from the class II
   force depends on the mixed neutron-proton densities and the generalized HFBTHO code is required to be used \cite{Sheikh2014}.

   \subsection{Results and Discussion}
   \label{sec:iso_dft_5}
\begin{table}[H]
\caption{Lagrangian parameters. \label{lambda}}
\newcolumntype{C}{>{\centering\arraybackslash}X}
\begin{tabularx}{\textwidth}{|C|C|C|C|C|C|}
\hline	
\multirow{2}{*}{$A$}&\multirow{2}{*}{Interaction}& \multicolumn{2}{C|}{ISB=0} &\multicolumn{2}{C|}{ISB=1}\\ \cline{3-6} 
&& $\lambda^\prime$ &$\lambda_{\textrm{off}}$&$\lambda^\prime$ &$\lambda_{\textrm{off}}$\\
\hline									
\multirow{2}{*}{48}&Without Coulomb&9.799936&0&10.1009095&0\\\cline{2-6}
&With Coulomb&11.078167&-8.064794&11.2299655&-8.4255485\\\hline
\multirow{2}{*}{78}&Without Coulomb&21.512809&0&21.9512975&0\\\cline{2-6}
&With Coulomb&23.7547175&-10.9167065&24.001203&-11.33256	\\			
\hline
\end{tabularx}
\end{table}
\begin{figure}[H]
 \centerline{\includegraphics[trim=0cm 0cm 0cm
0cm,width=0.9\textwidth,height=0.7\textwidth,clip]{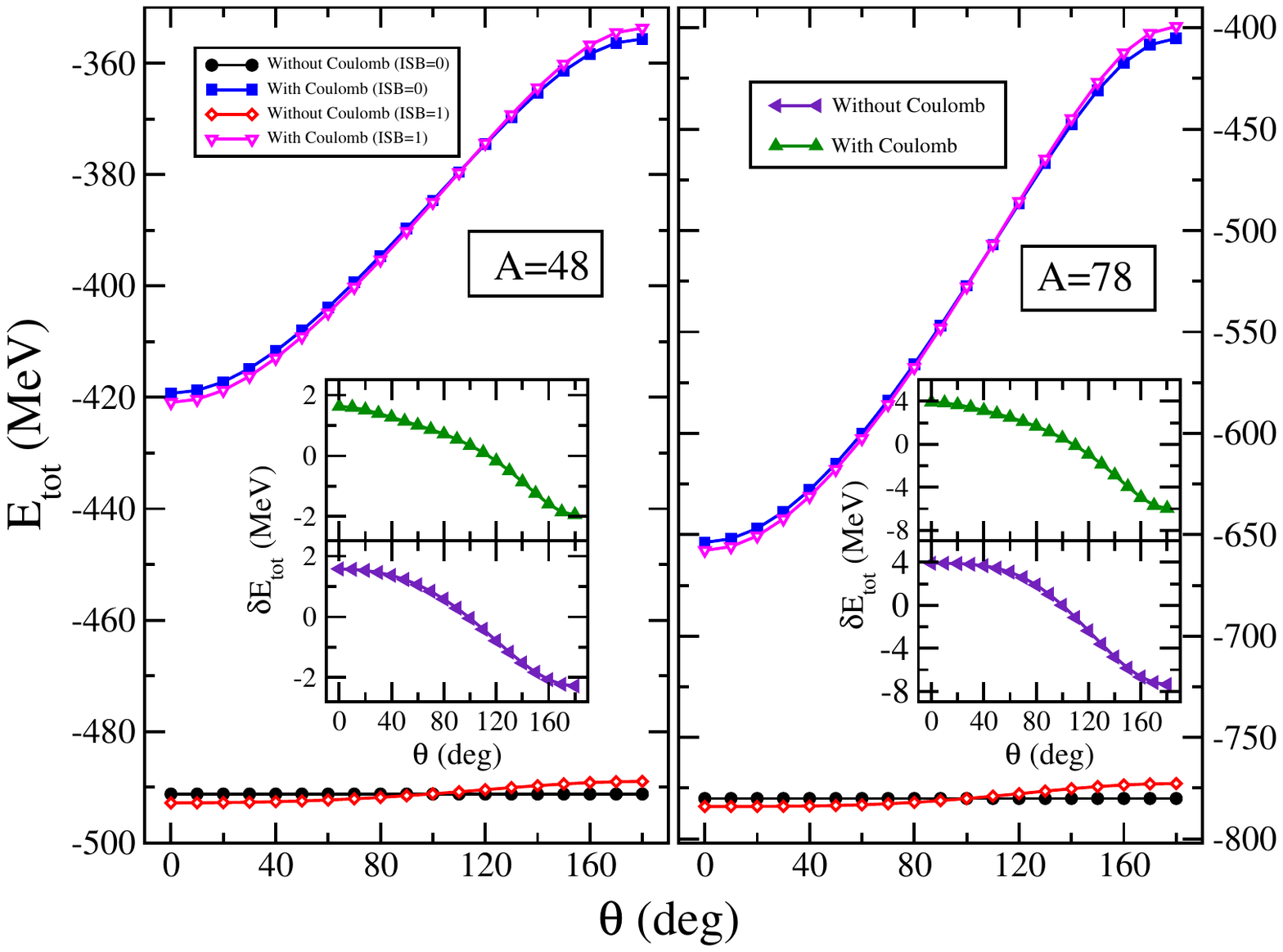}} \caption{(Color
online) Total HF energy of the $A=48$ and $78$ for the isospin states of $T=4$ and $11$, respectively. 
  }
\label{4878energy}
\end{figure}
   We have performed numerical calculations for a few isobaric chains of isotopes with the inclusion of the ISB terms in
   order to investigate the modifications on the nuclear properties. In particular, it would quite interesting to
   study the additional isospin mixing admitted by the ISB terms apart from the mixing with the Coulomb potential.
   The investigation has been undertaken for the isotopic chains of $A=48$ ($T=4$) and $A=78$ ($T=11$). These isobaric chains
   have already been studied in Ref.~\cite{Sheikh2014} and in the present work we shall explore the modifications
   on these already published results with the inclusion of the ISB terms. The Skyrme functional SkM$^{*}$ has been
   employed and the coupling constants $t_0^{II}=7 $ MeVfm$^3$ and  $t_0^{III}=-5.6 $ MeVfm$^3$ have been adopted from
   Ref.~\cite{BACZYK2018178}. As already stated, these parameters have been fitted to reproduce
   the MDEs and TDEs in lighter mass region and may not be optimum one's to study other properties or other regions.
   Nevertheless, the purpose of the present work is to have an overall picture of the modifications on the properties
   and these parameters will suffice.

   The Lagrangian parameters used in the isocranking model are listed in Table (\ref{lambda}) for the studied isobaric chains.
   In Fig.~(\ref{4878energy}), the total energy as a function of the isocranking angle, $\theta$ are displayed for the isobaric
   chains of $A=48$ (left panel) and 78 (right panel). The calculations have been carried out for four sets $:$ 
   (i) without Coulomb and without ISB , (ii) without Coulomb but with ISB , (iii) with Coulomb but without ISB
   and (iv) with Coulomb and also with ISB terms. The total energy without Coulomb and without ISB is independent
   of the $\theta$, which means that it has the same value for all the isobars in a given chain. This is expected as nuclear
   DFT of Eq.~(\ref{VIIEDF1}) is isoscalar and changing neutrons into protons will not alter the nuclear properties. There would be some
   spurious isospin mixing due to the mean-field approximation, but it is evident from the calculations that this is not
   very significant.
\begin{figure}[H]
 \centerline{\includegraphics[trim=0cm 0cm 0cm
0cm,width=0.9\textwidth,height=0.6\textwidth,clip]{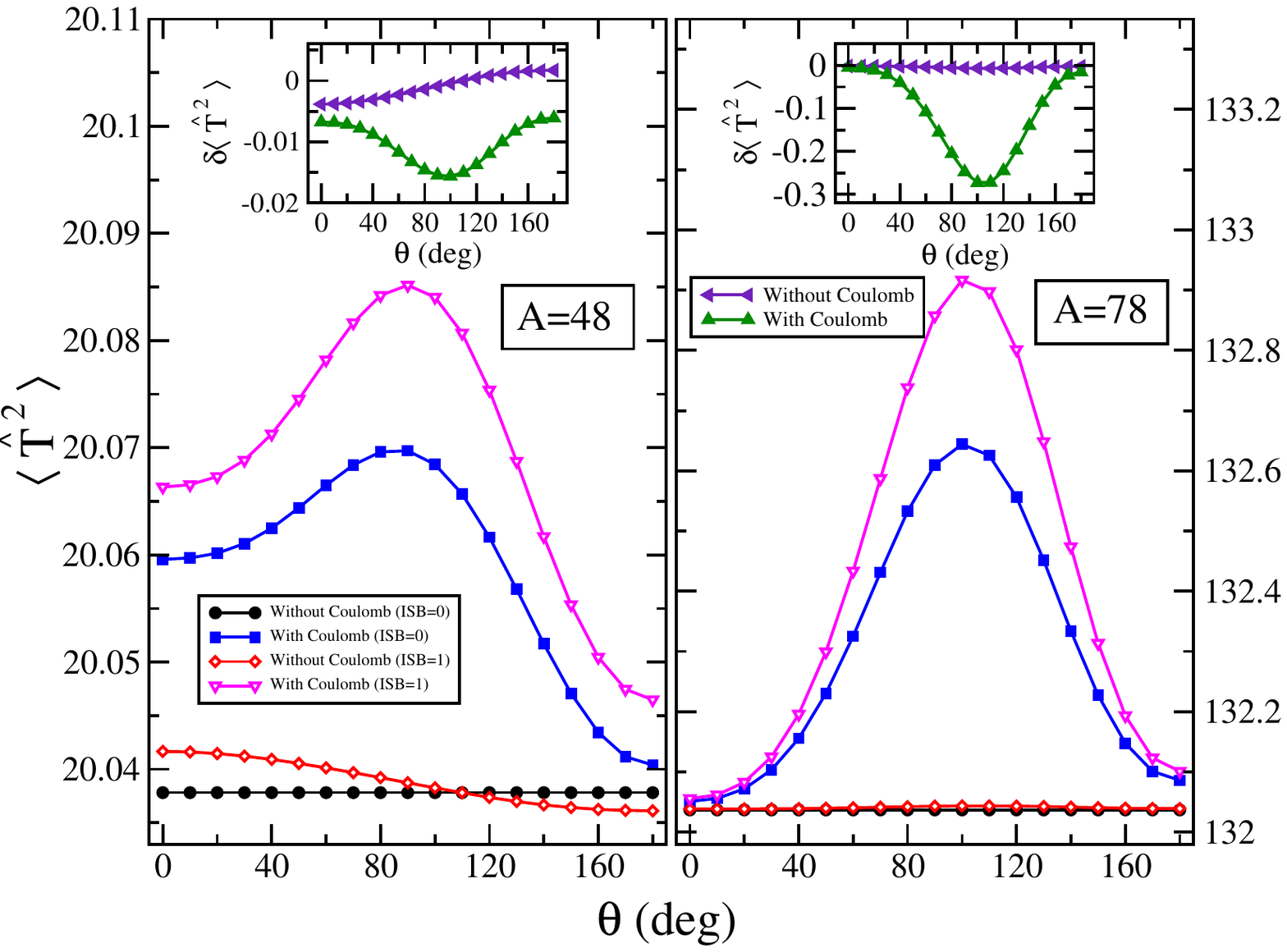}} \caption{(Color
online) $\langle \widehat{T}^2\rangle$ of the $A=48$ and $78$ for the isospin states of $T= 4$ and $11$. 
  }
\label{4878T2}
\end{figure}
\begin{figure}[H]
 \centerline{\includegraphics[trim=0cm 0cm 0cm
0cm,width=0.9\textwidth,height=0.6\textwidth,clip]{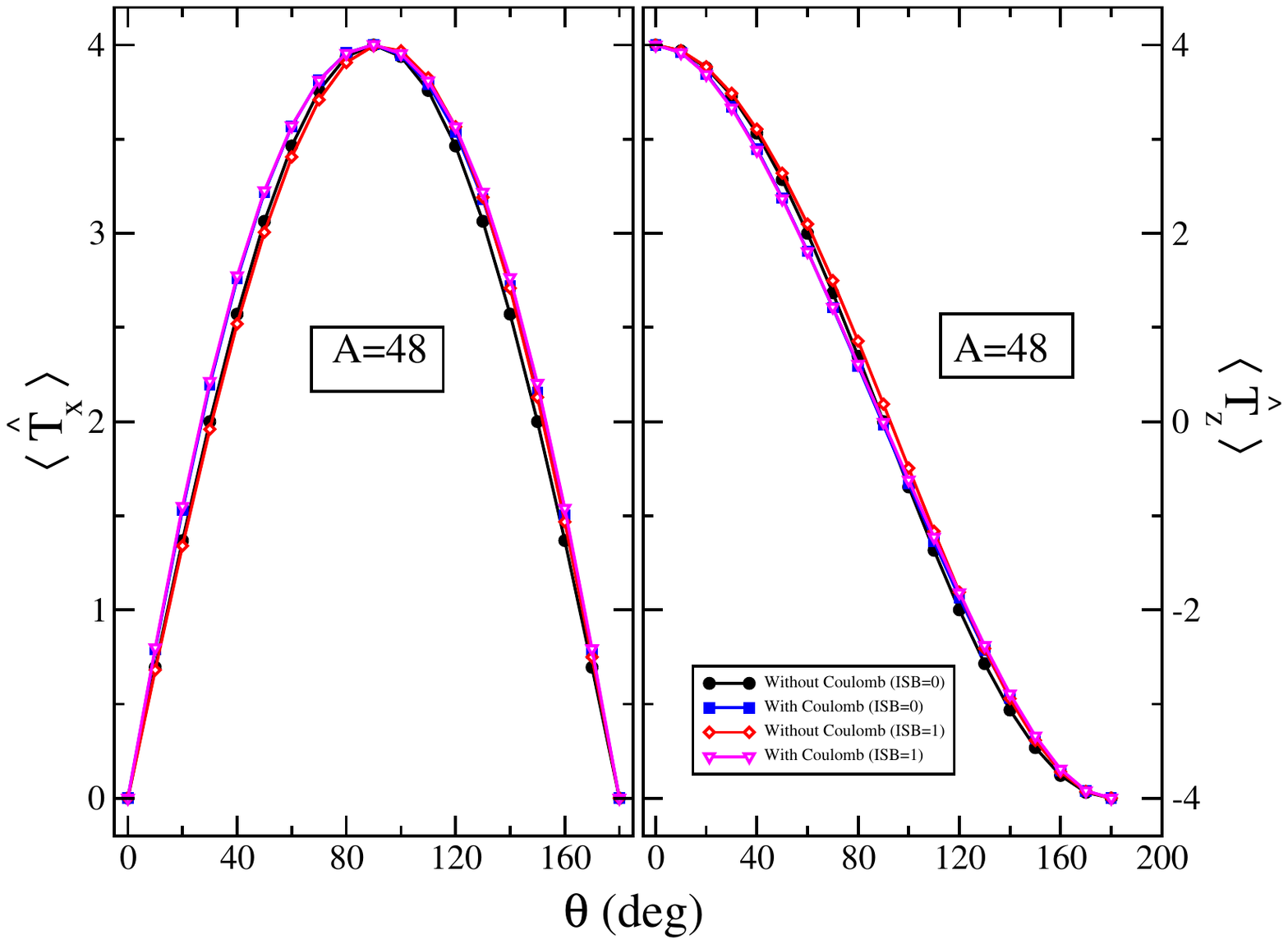}} \caption{(Color
online) Expectation values $\langle \widehat{T}_x\rangle$ and $\langle \widehat{T}_z\rangle$ of the $A=48$ for the isospin state $T=4$. 
  }
\label{48txtz}
\end{figure}

   Total energy calculated with ISB terms but without Coulomb is lower in energy before $\theta=90^\circ$ and above
   this angle, it is higher. This means that contribution before $\theta=90^\circ$ is attractive and above
   it has a repulsive behaviour. This can easily understood by noting that the ISB term of class II has "$t_z$"
   dependence and since its coupling constant is negative, the contribution of this term will be attractive for $N > Z$,
   zero for $N = Z$ and negative for $N < Z$. This is the behaviour noted for the total energy in Fig.~(\ref{4878energy})
   with the inclusion of the ISB terms. The contribution of ISB terms is about 2 MeV for the $A=48$ isobars
   of $^{48}$Ni and $^{48}$Ca, and is about 4 and 8 MeV for the $A=78$ isobars of $^{78}$Ni and $^{78}$Sn, respectively.

\begin{figure}[H]
 \centerline{\includegraphics[trim=0cm 0cm 0cm
0cm,width=0.9\textwidth,height=0.7\textwidth,clip]{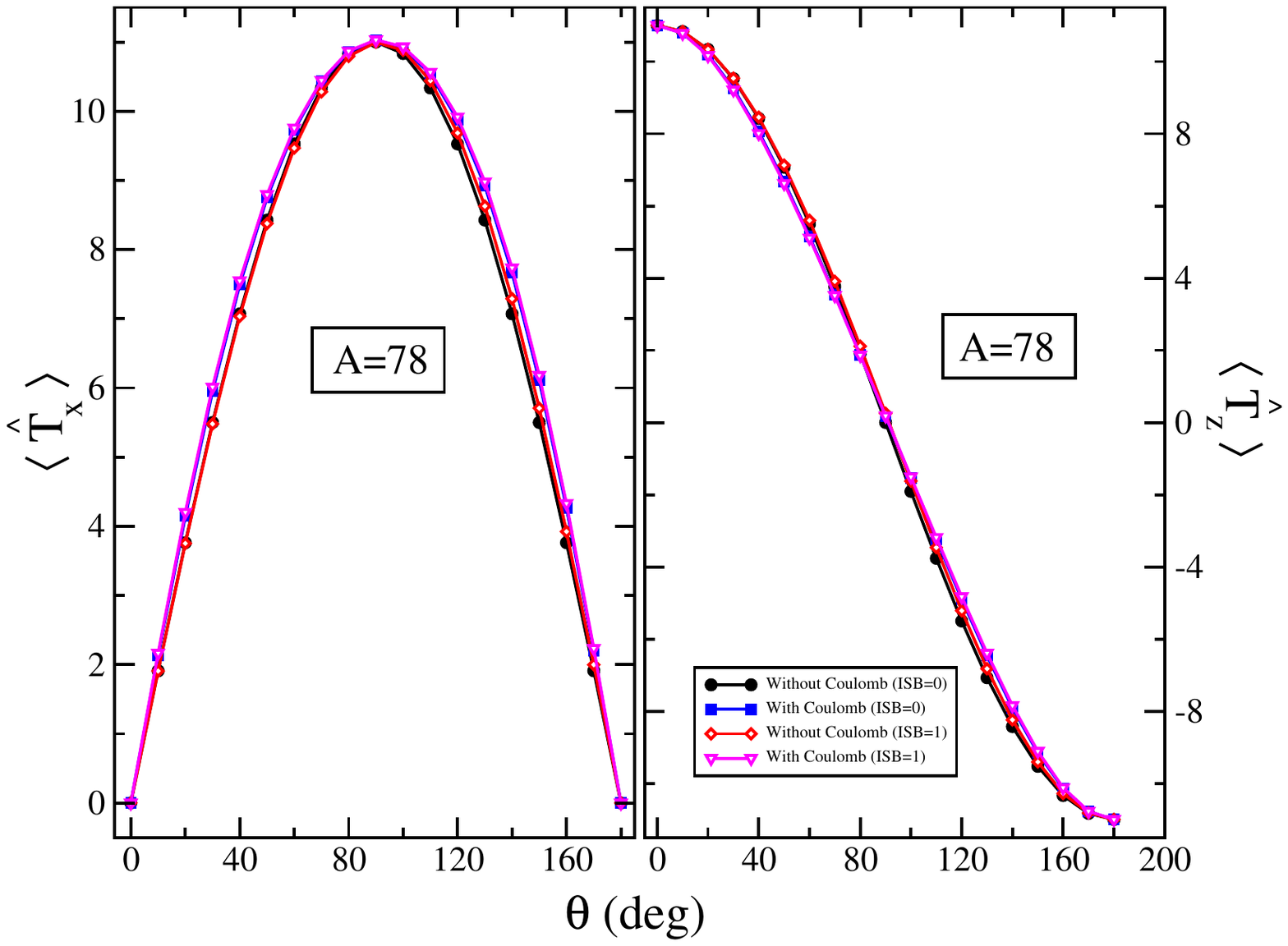}} \caption{(Color
online) Expectation values $\langle \widehat{T}_x\rangle$ and $\langle \widehat{T}_z\rangle$ of the $A=78$ for the isospin state $T= 11$. 
  }
\label{78txtz}
\end{figure}

   The expectation value of ${\hat T}^2$ is displayed in Fig.~(\ref{4878T2}) for the two isobaric chains. In the absence
   of the Coulomb potential and ISB, this value should be exactly equal to 20 (132), but this value
   is 20.03 (132.04) for $A=48$ (78) isobars. This small deviation is caused by the spurious
   isospin mixing of the mean-field approximation.
   In the presence of the ISB terms but without Coulomb, $\langle{\widehat T}^2\rangle$ is slightly higher than without ISB
   before $\theta=90^\circ$ and after this angle, it becomes lower. In the presence of the Coulomb potential,
   the results are different with $\langle{\widehat T}^2\rangle$ showing the maximum deviation around $\theta=90^\circ$, i.e.,
   for $N \sim Z$ system. The inclusion of the ISB terms, $\langle{\widehat T}^2\rangle$ depicts further deviations.
   The expectation values of $\langle T_x\rangle$ and $\langle T_z\rangle$ are depicted in Figs. (\ref{48txtz}) and (\ref{78txtz}) for the isobars of
   A=48 and 78, respectively. It is evident from the figures that there are minor deviations with and without
   ISB terms. In particular for the case without Coulomb interaction, the inclusion of ISB term
   for $\theta=90^\circ$ has no impact on $\langle J_x\rangle$ since for this angle, there is equal mixing of neutrons
   and protons.

\section{ Spherical shell model approach to isospin symmetry breaking }
\label{sec:iso_ssm}
The well-known phenomenological interactions like Cohen-Kurath interaction \cite{ckpot} for $p$-shell and USD interactions \cite{usd, usd1, usdb} for $sd$-shell are usually constructed as 
isospin-conserving interactions. While these interactions are widely successful in reproducing low-energy nuclear observables, the low-energy spectra obtained for different nuclei within the same 
isospin multiplet would be degenerate for such isospin-conversing interactions. The isospin symmetry is broken at the nuclear level due to the following three reasons :
\begin{enumerate}
\item Coulomb 
  interactions among protons, 
\item mass differences between proton and neutron, and
  \item the charge-dependent nature of nuclear forces.
\end{enumerate}

The isospin non-conserving interactions are required to study the difference in low-energy spectra between mirror pair nuclei whose proton ($Z$) and neutron ($N$) numbers having the 
same mass number ($A$) are exchanged. The mirror energy difference (MED) between the isospin analogue state (IAS) is a measure of the ISB in atomic nuclei, and it is defined as :

\begin{eqnarray*}
MED_I = E_I(T_z = -T) - E_I(T_z = +T),
\end{eqnarray*}
Here, $E_I$ are the excitation energies of analogous states with angular momentum $I$ in a mirror pair with $T_z$ = $\pm  T$.

To discuss isospin-related properties, we have performed calculations using the no-core shell model \cite{NCSM_r2} and the shell model for a few selected pairs of nuclei across the $sd$-shell region. Since the no-core shell model is computationally expensive for heavier nuclei, we consider only one $t_z$ = 1 mirror pair ($^{20}$Na - $^{20}$F) in the lower $sd$-shell region. Recently, $t_z$ = 1/2 mirror pair ($^{21}$Na - $^{21}$Ne) has been studied in Ref. \cite{Na_NCSM} within the no-core shell model formalism. In order to study the effect of ISB with the shell model, we have selected four pairs of nuclei in the mid $sd$-shell region corresponding to $t_z$ = $\pm$ 1/2 ($^{25}$Al/ $^{25}$Mg), $t_z$ = $\pm$ 1 ($^{26}$Si/ $^{26}$Mg), $t_z$ = $\pm$ 3/2 ($^{23}$Ne/ $^{23}$Al), and $t_z$ = $\pm$ 2 ($^{24}$Ne/ $^{24}$Si). We have calculated the low-energy spectra for the above-mentioned pairs of nuclei using the newly developed USDC \cite{usdc} interaction and compared them to the corresponding experimental data.

\subsection{No-core shell model results}
\begin{figure}[H]
    \centering
\includegraphics[scale = 0.55]{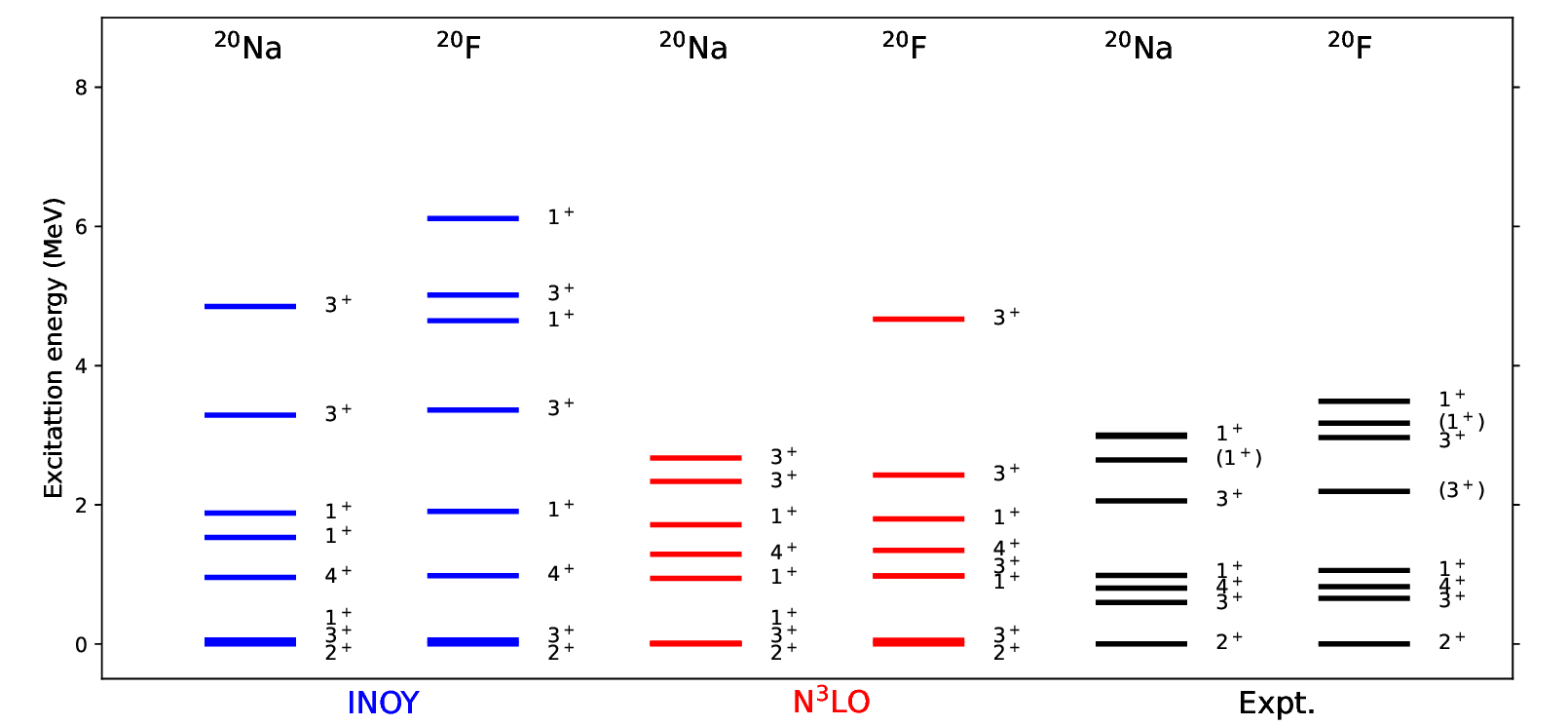}
\caption{\label{ncsm} Comparison of no-core shell model results with the experimental data for low-lying spectra of $^{20}$Na and $^{20}$F isotopes using INOY and N3LO interactions.}
\end{figure}

We have performed \textit{ab initio} no-core shell model calculations for $t_z$ = 1 mirror pair ($^{20}$Na - $^{20}$F) to study the effect of ISB. The low-energy spectra for these two nuclei are shown in Fig. \ref{ncsm}. We have used two $NN$ interactions, namely, the inside nonlocal outside Yukawa (INOY) \cite{inoy} and the chiral next-to-next-next-to-leading order (N3LO) \cite{n3lo} interactions. The INOY interaction is a phenomenological $NN$ interaction whose short-range part (r $<$ 1 fm) is modified by a controllable non-local part, and the long-range part (r $>$ 3 fm) is similar to the Yukawa tail of the Argonne V18 (AV18) \cite{av18} interaction. The N3LO interaction used in this work is a chiral interaction obtained from QCD using chiral perturbation theory ($\chi$PT).

The first step in the no-core shell model is to decide an optimum frequency for each interaction and it is obtained by plotting ground-state energies with harmonic oscillator ($\hbar\omega$) frequency for different model spaces controlled by the parameter, $N_{max}$. The no-core shell model results reported in this work correspond to $N_{max}$ = 4, and optimum frequencies 20 and 14 MeV, respectively, for INOY and N3LO interactions. From Fig. \ref{ncsm}, it can be seen that the ordering of the ground and first excited states for both nuclei are well reproduced by INOY and N3LO interactions. However, these two states are obtained to be degenerate up to $N_{max}$ = 4 model space calculation. Better results for the overall spectra can be expected for higher $N_{max}$ calculation. Experimentally, the ($1^+_2$) and $1^+_3$ states have large MEDs of -527 and -487 keV, respectively. The calculated MEDs using INOY (N3LO) are -3.115 (-0.852) and -4.234 (-2.956) MeV, respectively, for $1^+_2$ and $1^+_3$ states.

\begin{figure}[H]
    \includegraphics[scale=0.35]{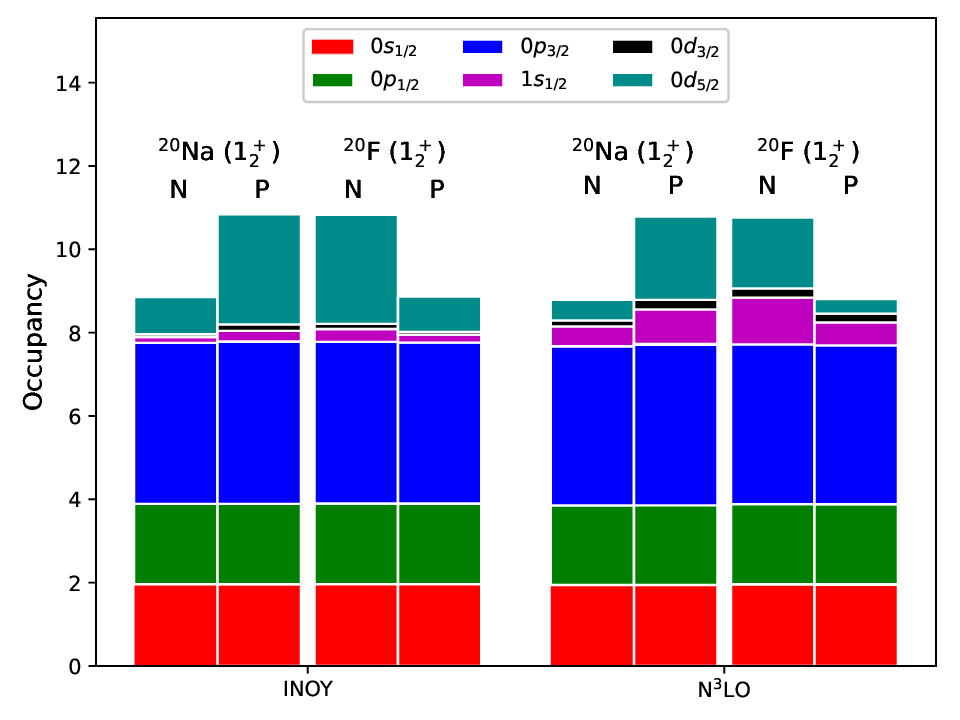}
    \includegraphics[scale=0.35]{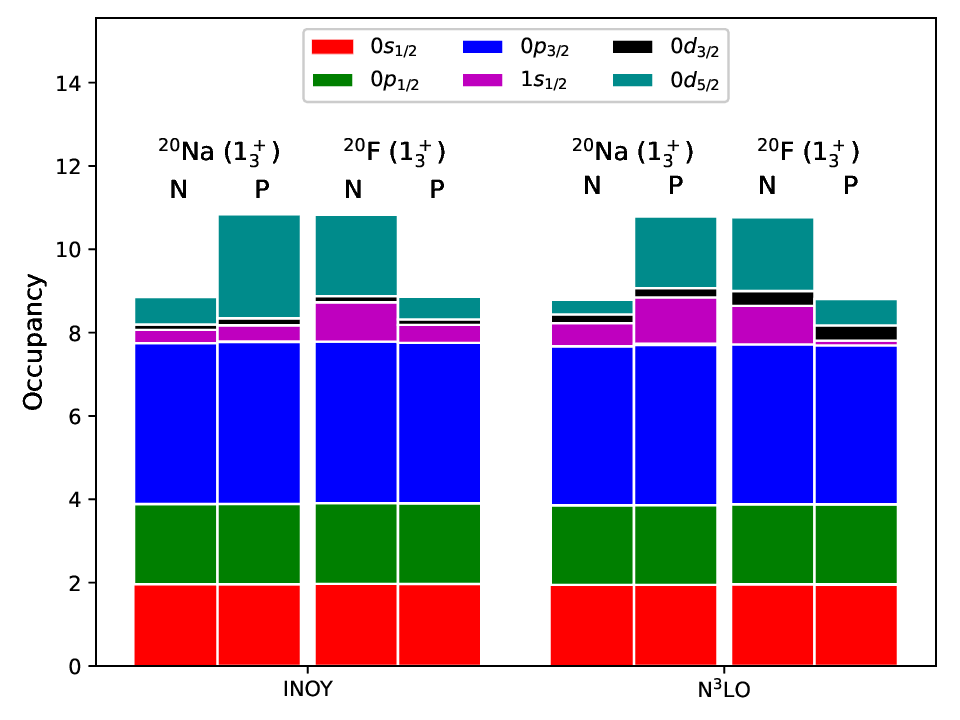}
    \caption{The neutron (N) and proton (P) occupancies of 1$^+_2$ and 1$^+_3$ states are shown for $^{20}$Na/$^{20}$F mirror pair.}
    \label{ncsm_occupancy}
\end{figure}

In Fig. \ref{ncsm_occupancy}, we have plotted the occupancies of single nucleon harmonic oscillator orbitals up to $N$ = 2 major shell for the 1$^+_2$ and 1$^+_3$ states. From the occupancy plots, it can be seen that the large MEDs in the NCSM results are due to the difference in occupancies of $\pi(1s_{1/2})$ of $^{20}$Na and $\nu(1s_{1/2})$ of $^{20}$F. Additionally, the difference in two body matrix elements (TBME) involving $1s_{1/2}$ orbitals for proton and neutron are also responsible for the large MEDs for these two states.

\subsection{Shell model results}
\begin{figure}[H]
    \centering
    \includegraphics[scale = 0.55]{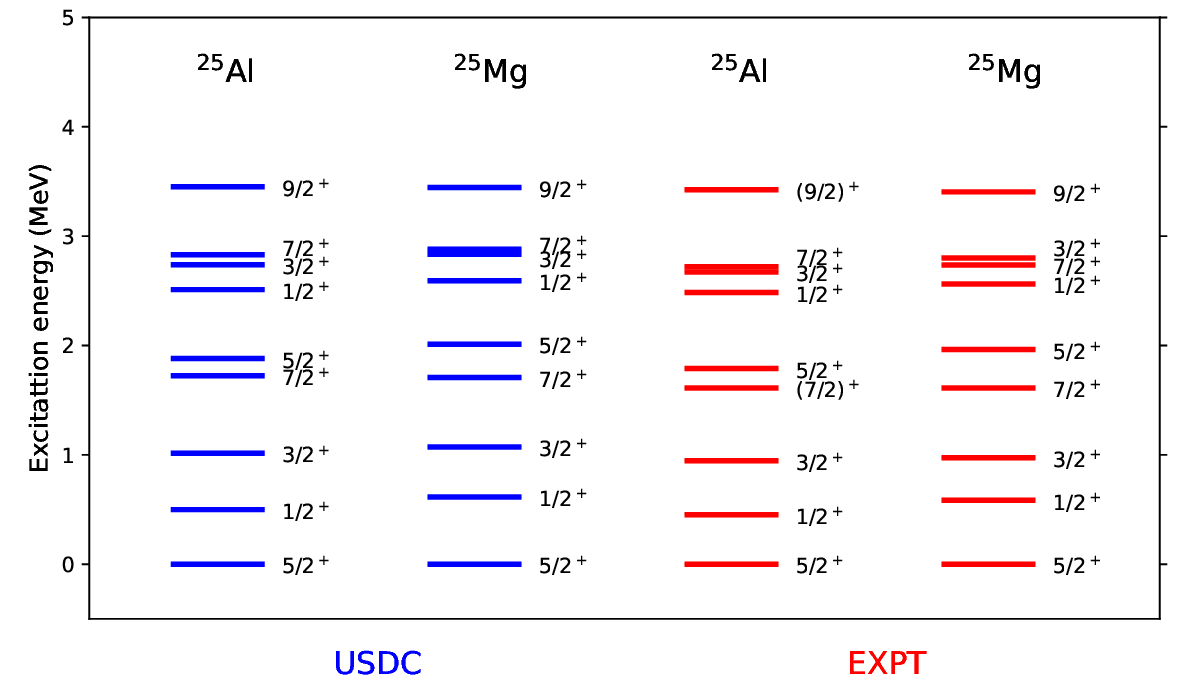}
\caption{Comparison of shell model results with the experimental data corresponding to low-lying states for  $t_z$ = $\pm$ 1/2 mirror pair ($^{25}$Al - $^{25}$Mg).}
\label{25Mg}
\end{figure}

\begin{figure}[H]
\centering
    \includegraphics[scale = 0.55]{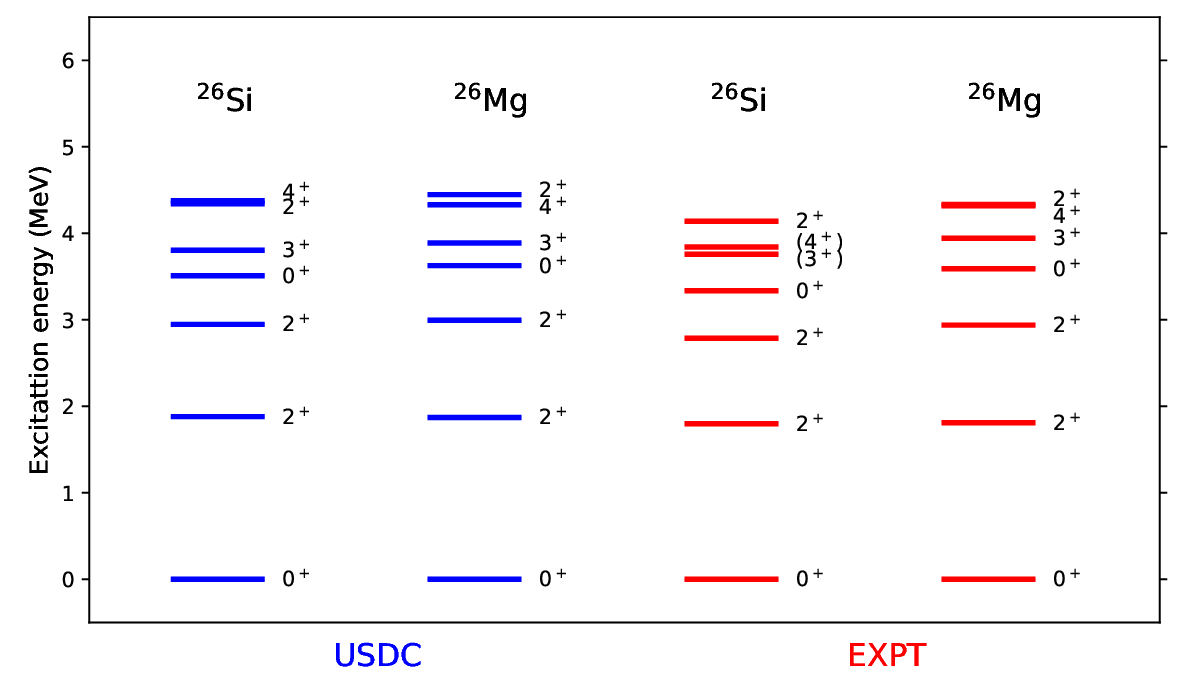}
\caption{Comparison of shell model results with the experimental data corresponding to low-lying states for $t_z$ = 1 mirror pair ($^{26}$Si - $^{26}$Mg).}
\label{26Mg}
\end{figure}

\begin{figure}[H]
\centering
    \includegraphics[scale = 0.55]{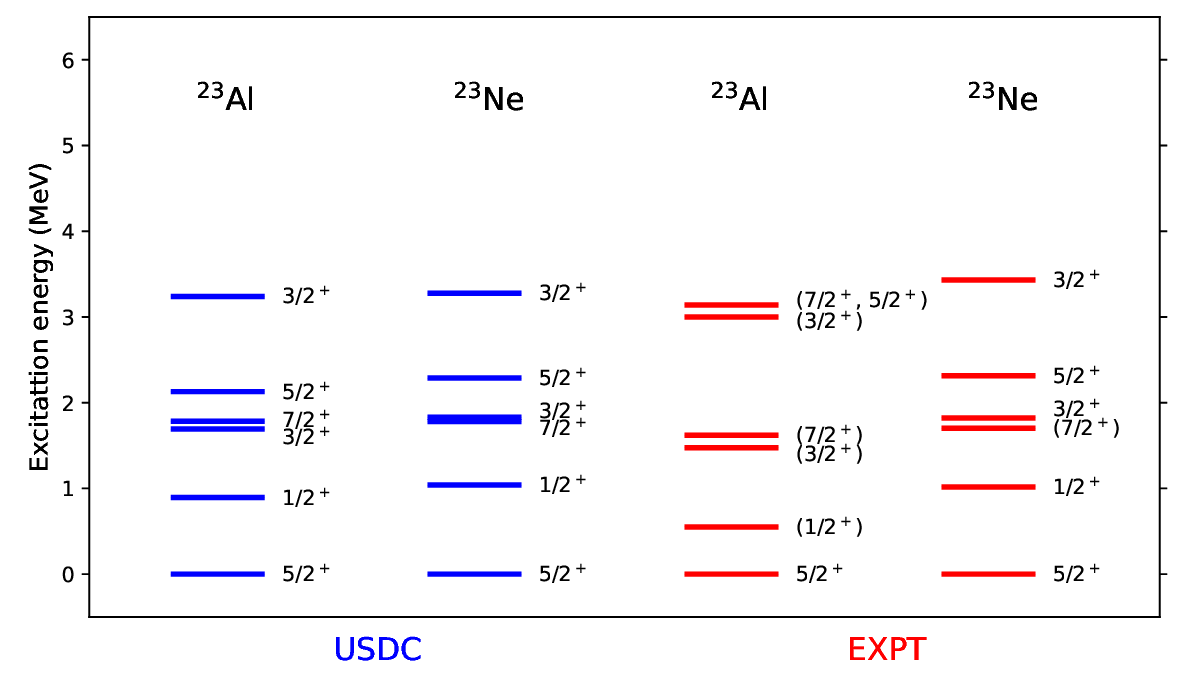}
\caption{Comparison of shell model results with the experimental data corresponding to low-lying states for  $t_z$ = 3/2 mirror pair ($^{23}$Al- $^{23}$Ne).}
\label{23Ne}
\end{figure}

\begin{figure}[H]
\centering
    \includegraphics[scale = 0.55]{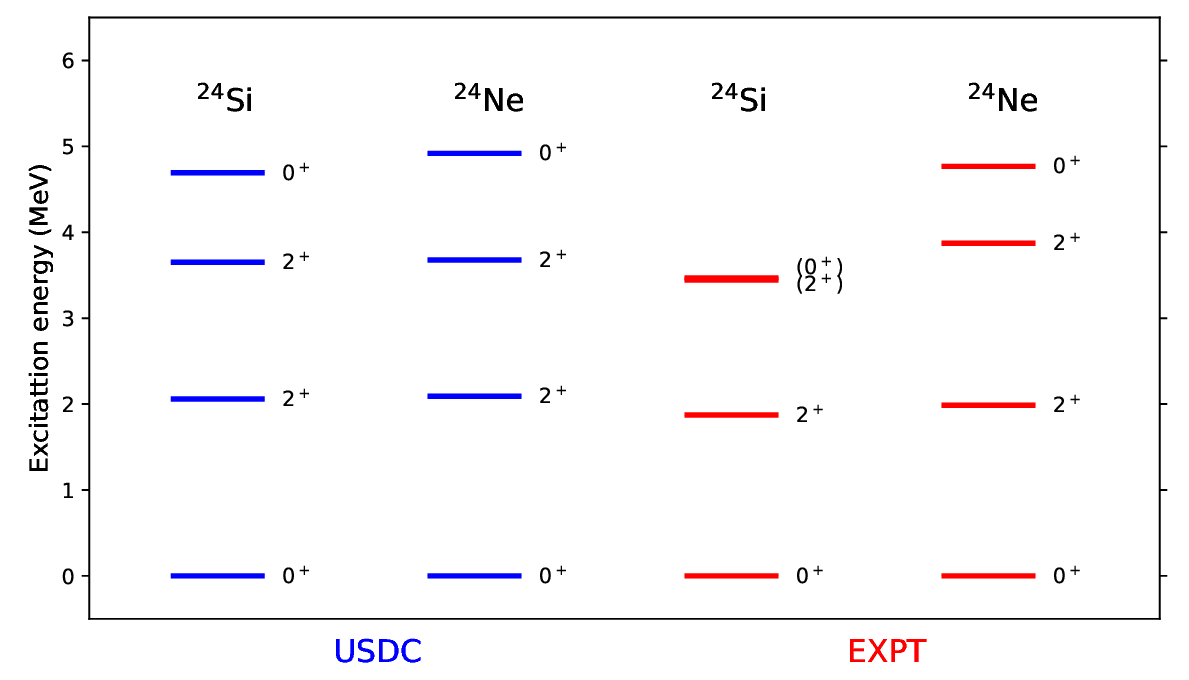}
\caption{Comparison of shell model results with the experimental data corresponding to low-lying states for $t_z$ = $\pm$ 2 mirror pair ($^{24}$Si- $^{24}$Ne).}
\label{24Ne}
\end{figure}

\begin{figure}[H]
    \centering
    \includegraphics[scale = 0.65]{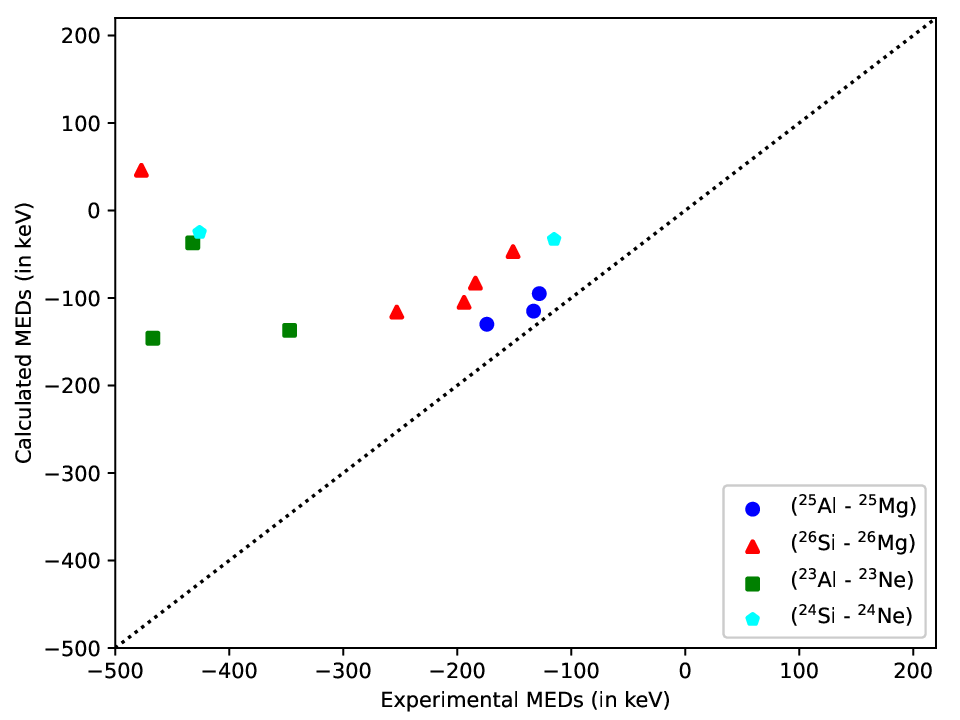}
    \caption{The experimental MEDs are compared to the calculated ones using USDC interaction for a few states whose observed MEDs are more than 100 keV. }
    \label{MED_scatter}
\end{figure}

\begin{figure}[H]
    \centering
    \includegraphics[scale = 0.65]{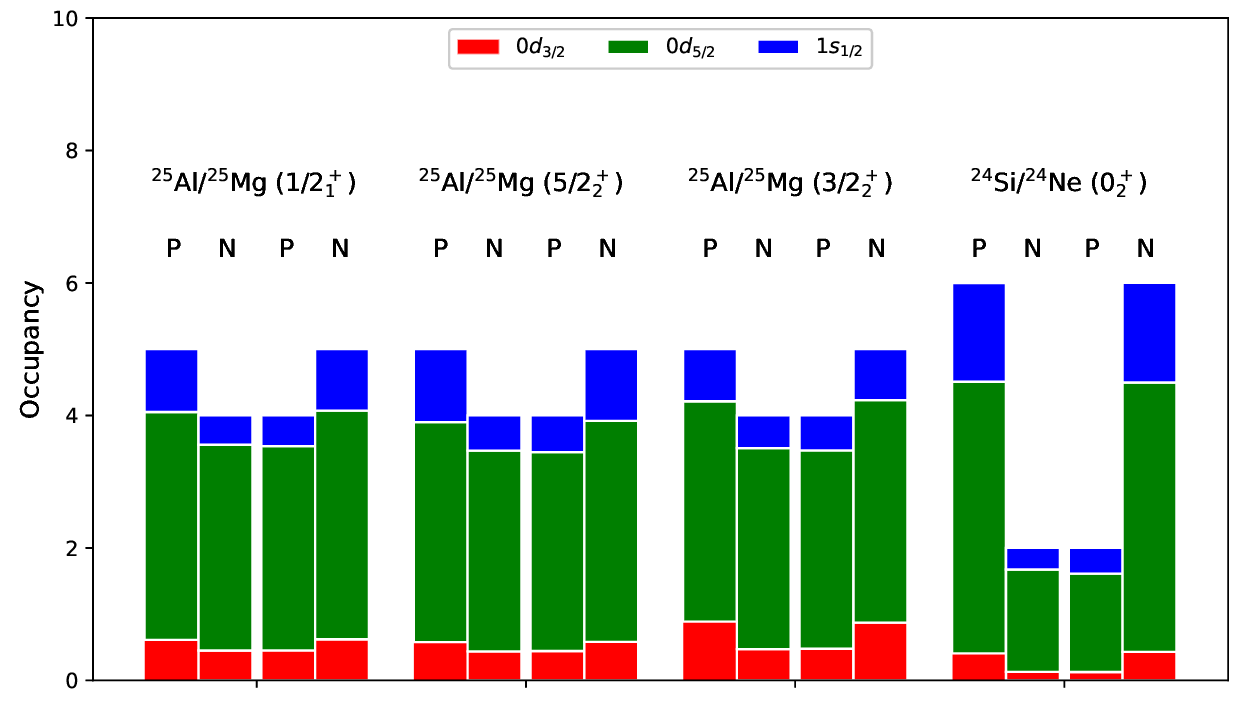}
    \caption{Comparison of calculated proton (P) and neutron (N) occupancies of 1/2$^+_1$, 5/2$^+_2$, and 3/2$^+_2$ states of $^{25}$Al/$^{25}$Mg pair and 0$^+_2$ state of $^{24}$Si/$^{24}$Ne mirror pair using shell-model with USDC interaction.}
    \label{MED_occ}
\end{figure}

To study the effect of ISB within the shell model formalism, we have used the isospin non-conserving USDC interaction. This interaction is fitted to 854 states of 117 nuclei across the $sd$-shell. The USDC interaction is constrained by a renormalized G-matrix Hamiltonian, and an analytic Coulomb part is added along with a 2.2 \% increase in the proton-neutron $(T = 1)$ TBMEs. The full USDC Hamiltonian is provided in proton-neutron formalism comprising of three parts : 1) the isospin-conserving part, 2) the isospin-breaking part, and 3) the Coulomb part. We applied this newly developed interaction to calculate MEDs for $T_z$ = $\pm$ 1/2 ($^{25}$Al/ $^{25}$Mg), $T_z$ = $\pm$ 1 ($^{26}$Si/ $^{26}$Mg), $T_z$ = $\pm$ 3/2 ($^{23}$Ne/ $^{23}$Al), and $T_z$ = $\pm$ 2 ($^{24}$Ne/ $^{24}$Si) mirror pairs in the mid $sd$-shell region and the results are shown in Figs. \ref{25Mg} - \ref{MED_occ}.

Fig. \ref{25Mg} shows the comparison between the calculated and experimental low-energy spectra of $t_z$ = $\pm$ 1/2 mirror pair ($^{25}$Al - $^{25}$Mg). The shell model results for USDC interaction are in good agreement with the experimental low-energy spectra, except the ordering of 3/2$^+_2$ and 7/2$^+_2$ are reversed for $^{25}$Mg compared to the experimental data. Out of the nine low-lying states considered in this work, three states, namely 1/2$^+_1$, 5/2$^+_2$, and 3/2$^+_2$ show MEDs greater than -100 keV. The experimental (calculated) MEDs of these states are -133 (-115), -175 (-130), and -128 (-95) keV, respectively. The three blue circles in Fig. \ref{MED_scatter} represent these three states.

A comparison between the calculated and experimental spectra consisting of seven low-lying states of $T_z$ = $\pm$ 1 mirror pair ($^{26}$Si - $^{26}$Mg) is shown in Fig. \ref{26Mg}. The shell model results are in good agreement with experimental spectra except the ordering of 2$^+_3$ and 4$^+_1$ states are reversed for the case of $^{26}$Si. Out of these seven low-lying states, five states show more than 100 keV MEDs. The experimental (calculated) MEDs of 2$^+_2$, 0$^+_2$ and 2$^+_3$ states, which have confirmed spin-parity in both nuclei, are -151 (-47), -252 (-116), and -193 (-105) keV, respectively. A large MED of 477 keV is observed experimentally between the (4$^+$) state of $^{26}$Si and 4$^+$ state of $^{26}$Mg. The calculated MED for this state is 46 keV compared to the experimental MED of 477 keV. Further investigation of this state would be interesting using different charge-dependent interactions. The five red triangles in Fig. \ref{MED_scatter} represent those five states with MEDs more than -100 keV.

The low-energy spectra of $T_z$ = $\pm$ 3/2 mirror pair ($^{23}$Al - $^{23}$Ne) is shown in Fig. \ref{23Ne}. The calculated results for six low-lying states are in good agreement with the experimental data. Out of these states, 1/2$^+_1$, 3/2$^+_1$ and 3/2$^+_2$ show large MEDs of -467, -347, and -432 keV, respectively. The calculated results are -146, -137, and -37 keV, respectively, corresponding to those three states. A comparison of calculated and experimental MEDs for these three states is shown as three green squares in Fig. \ref{MED_scatter}. 

Lastly, Fig. \ref{24Ne} shows the comparison between the calculated and experimental low-energy spectra of $T_z$ = $\pm$ 2 mirror pair ($^{24}$Si - $^{24}$Ne). All three excited states for this pair of nuclei show large MEDs : 2$^+_1$ (-110 keV), 2$^+_2$ (-426 keV), and 0$^+_2$ (-1.298 MeV). The shell results for MEDS of these states are too small compared to the experimental data. The large MEDs in the 0$^+_2$ state can be explained by considering the Thomas-Ehrman shift \cite{TES1}. The two cyan pentagons in Fig. \ref{MED_scatter} represent the other two states except for 0$^+_2$, which show larger than 100 keV MED. 

In Fig. \ref{MED_occ}, we have plotted the proton (P) and neutron (N) occupancies of 1/2$^+_1$, 5/2$^+_2$, and 3/2$^+_2$ states of $^{25}$Al/$^{25}$Mg pair along with the 0$^+_2$ state of $^{24}$Si/$^{24}$Ne mirror pair. From the figure, it can be seen that the proton occupancies of the proton-rich partner can be compared to the neutron occupancies of the neutron-rich partner and vice-versa, and the MEDs are solely due to the difference in $pp$, $nn$, and $np$ part of the USDC interaction. 

\section{ Summary and conclusions }
\label{sec:iso_summary}

The concept of isospin symmetry, originally introduced by Heisenberg \cite{Heisenberg1989} and Wigner \cite{Ewigner1937}, has played
a pivotal role in the development of particle and nuclear physics models. It contributed to the development of eight-fold way
for the classification of elementary particles. The utility of the isospin symmetry
in nuclear physics was demonstrated very early by Wigner \cite{Ewigner1957}. In this work,
the nucleon-nucleon interaction was considered to be isoscalar and the only isospin violating term
was the Coulomb interaction, which was treated using the perturbation theory. It was shown that mass excess
of the isobaric analogue states can be expressed as a quadratic polynomial in $T_Z$ with the coefficients depending on
the isospin quantum number. This is what is referred to as the IMME and it known to describe the masses of the
analogue states quite accurately. As a matter of fact, this model has been found to be quite useful to predict the nuclear masses
of the analogue states for which it is difficult to perform the measurements \cite{MacCormick2014}. The
masses of the isobaric analogue states are needed in modelling the stellar evolution.

The isospin symmetry has provided a useful framework for the
classification of the nucleon-nucleon interaction. The nuclear potential is expressed in terms of isoscalar, isovector and isotensor
components, what is referred to as Henley-Miller classification \cite{Henleybook1969,Henleybook1979,Miller1990}. The dominant part of the nuclear interaction is the
isoscalar component and the isospin symmetry is weakly broken by isovector and isotensor components. This is evident from the
empirical data available from nucleon-nucleon scattering. ISB has two manifestations : CSB and CIB. CSB breaking is a subgroup of the CIB as it involves
a special rotational invariance in the isospin space. In the Henley-Miller classification scheme, class I force
is scalar in isospin space, Class II force maintains
charge symmetry, but breaks charge independence and class III and IV break both charge symmetry and charge independence.

The \textit{ab initio} no-core shell model and shell model are two powerful approaches to study the small difference in isobaric analogue states due to ISB. While the no-core shell model results overpredict the corresponding MEDs, the shell model results with USDC interaction underpredict the experimental MEDs. Better results for the no-core shell model can be expected for higher $N_{max}$ calculations, and some improved results from shell model calculations can be obtained by using USDCm and USDIm interactions \cite{usdc} where the Coulomb terms are further constrained. 

In an attempt to describe MDEs and TDEs using the nuclear DFT approach, the ISB terms of class II and III have
been introduced in the nuclear density functional \cite{BACZYK2018178}. The conventional Skyrme DFT is unable to illustrate these properties,
and, in particular, the staggering observed in TDEs is completely absent in the DFT calculations. It has been demonstrated \cite{BACZYK2018178}
that with the inclusion of the ISB terms in the Skyrme DFT, staggering observed in the data is reproduced. The two parameters
have been fitted to the masses of the lighter mass nuclei independently without tweaking the original coupling constants
of the Skyrme functional. We have investigated the isobaric chains of isotopes using these newly introduced ISB terms in the
energy functional. The study has been performed using the isospin cranking approach, which is an approximation to the
isospin projection. It has been shown that these terms are attractive in nature for $ N > Z$ and become repulsive for $Z < N$.
For the $N=Z$ system, there is no contribution of the ISB term to the energy.

For an accurate treatment of ISB terms in DFT, isospin projection needs to be performed. The ISB term of class II involves
neutron-proton mixed densities and the proton and neutron particle numbers are not well defined in the HF approach. Therefore,
three-dimensional isospin projection is required to be carried out. In the absence of the ISB terms, one-dimensional
isospin projection  has been performed in Skyrme DFT approach and it has been shown that this particular projection
is free from the divergence problem that particle-number and angular-momentum projection is beset with. For this
special case, the singularity problem has been analyzed analytically and it would be interesting to investigate
whether the three-dimensional isospin projection is also free from the divergence problem.

The coupling constants of the ISB terms have been separately fitted to the masses in the lighter mass
region. Although the contribution of ISB terms is quite small, nevertheless, in an accurate treatment the parameters
of the Skyrme functional need to be refitted by also involving the coupling constants of the ISB terms in the
fitting protocol. Further, the ISB terms introduced in the energy functional have been adopted from the shell model
perespective. It is very essential to derive  the ISB terms of the nuclear potential from the \textit{ab initio}
methods, for instance, following the chiral effective field theory approach.

\acknowledgments{We would like to express our gratitude to the High-Performance Computing (HPC) facility through IUST Cloud at Islamic University of Science and Technology, Awantipora, J$\&$K, India and the  Promotion of University Research and Scientific Excellence (PURSE) under the project No. SR/PURSE/2022/121(C). The computational resources provided by the HPC facility, supported by the PURSE project, have been used to perform some of the simulations presented in this manuscript. The authors are also thankful to Prof. Nobuo Hinohara for useful discussions.}

\begin{appendices}
  \renewcommand{\theequation}{\Alph{section}\arabic{equation}}
 \setcounter{equation}{0} 
\section{Lowest order Perturbation}
\label{appendix1}
\begin{figure}[htb]
 \centerline{\includegraphics[trim=0cm 0cm 0cm
0cm,width=0.5\textwidth,clip]{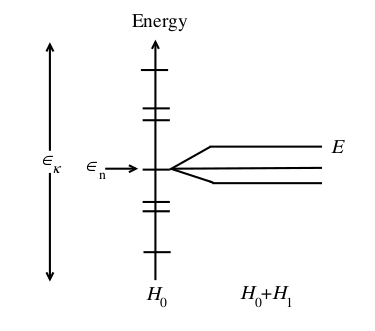}} \caption{(Color
online) The energy levels $E_k$ of the unperturbed system $\widehat{H}_0$ are shown on a vertical axis. These levels are degenerate in general. One of these, $\epsilon_n$, is selected out for study. When the perturbation $\widehat{H}_1$ is turned on, level $\epsilon_n$ may split and shift. The energy level $E$ of the perturbed system $\widehat{H}_0+\lambda \widehat{H}_1$ is one that grows out of $\epsilon_n$ as the perturbation is turned on. 
  }
\label{Pertth1}
\end{figure}

\begin{figure}[htb]
 \centerline{\includegraphics[trim=0cm 0cm 0cm
0cm,width=0.5\textwidth,clip]{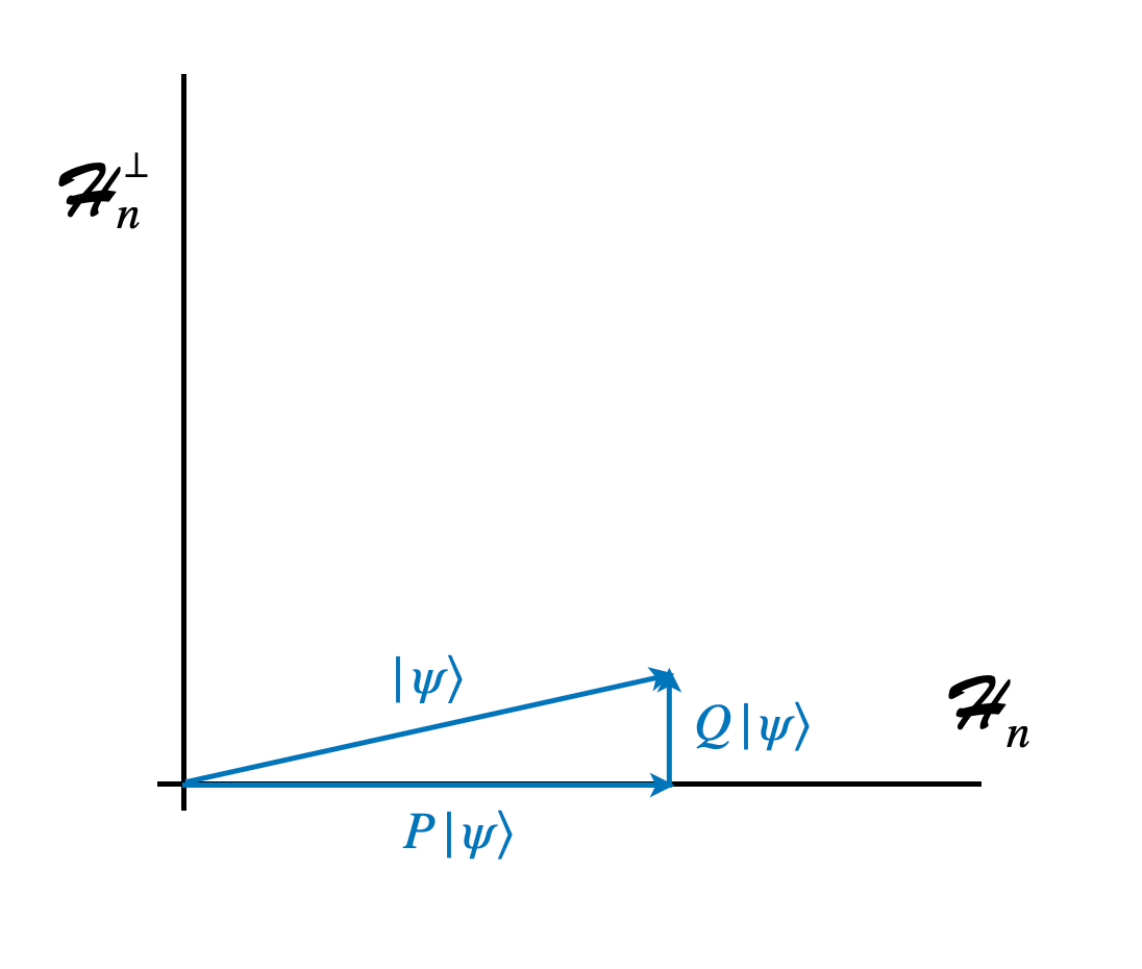}} \caption{(Color
online) Subs[pace $\widehat{\mathcal{H}}_n$ is unperturbed eigenspace for level $E_n$, and $\widehat{\mathcal{H}}_n^\perp$ is the orthogonal subspace. Projectors $\widehat{P}$ and $\widehat{Q}$ project onto  $\widehat{\mathcal{H}}_n$ and $\widehat{\mathcal{H}}_n^\perp$, respectively. The exact eigenstate lies approximately in $\widehat{\mathcal{H}}_n$, with a small correction $\widehat{Q}|\psi\rangle$ orthogonal to this subspace. 
  }
\label{Pertth2}
\end{figure}

Consider the perturbed Hamiltonian
 \begin{eqnarray}
   \widehat{H}=\widehat{H}_0+\lambda \widehat{H}_1~,
 \end{eqnarray}
 where, $\lambda$ denotes the formal expansion parameter with values between 0 and 1. When $\lambda\neq 0$, there may be splitting and shifting of an unperturbed energy level. Let $E$ be one of the exact energy  
 levels of $H$ as shown in Fig. \ref{Pertth1}, corresponding to the the exact energy eigenket $|\psi \rangle$, so that
 \begin{eqnarray}
   \widehat{H}|\psi\rangle=\widehat{H}_0+\lambda \widehat{H}_1|\psi \rangle=E|\psi \rangle~.
   \label{HPSI}
 \end{eqnarray}
We assume that the above exact eigenket $|\psi\rangle$ arises as a result of the perturbation from the eigenket $|\psi)$ (which may be degenerate) of $\widehat{H}_0$ with eigen energy $\epsilon_n$. Understanding the perturbation 
problem in terms of Hilbert space geometry is helpful. We divide the Hilbert space, spanned by the complete eigen set of $\widehat{H}_0$ (assumed to be orthonormal)  into the $\mathcal{H}_n$ subspace, which is spanned by the eigen set $|\psi)=|n\alpha)$ of $\widehat{H}_0$ 
with energy $\epsilon_n$ and its  orthogonal counterpart $\mathcal{H}^{\perp}_n$, which is spanned by the rest of the eigen set $|l\alpha)$, $l\neq n$, of $\widehat{H}_0$. Here $\alpha$ is the set of all those quantum numbers that are 
necessary to designate the quantum state uniquely but do not determine the energy of the state and hence dictate the degeneracy of the system. 
The exact eigenket $|\psi\rangle$ probably resides nearly in $\mathcal{H}_n$ with only a minor component orthogonal to $\mathcal{H}_n$ because the perturbation is small, as seen in Fig. \ref{Pertth2}. The components 
of $|\psi\rangle$ parallel and perpendicular to $\mathcal{H}_n$ can be represented in terms of the projector $\widehat{P}$ onto the subspace $\mathcal{H}_n$ and the orthogonal projector $\widehat{Q}$, which are defined by
 \begin{eqnarray}
   \widehat{P}=\sum_\alpha|n\alpha)(n\alpha | ~~~~~\rm{with}~~~~~ \widehat{P}^2=\widehat{P}~,
 \end{eqnarray}
 and
 \begin{eqnarray}
  \widehat{Q}=\sum_{l\alpha}|l\alpha)( l\alpha|  ~~~~~\rm{with}~~~~~\widehat{Q}^2=\widehat{Q}~.
 \end{eqnarray}
 These projectors also satisfy the relations
 \begin{eqnarray}
   \widehat{P}+\widehat{Q}=I~,~~~~~\widehat{P}\widehat{Q}=\widehat{Q}\widehat{P}=0~.
 \end{eqnarray}
 Also they commute with $\widehat{H}_0$
 \begin{eqnarray}
   [\widehat{P}, \widehat{H}_0]=[\widehat{Q}, \widehat{H}_0]=0~.
 \end{eqnarray}
 Eq. (\ref{HPSI}) can be written as
 \begin{eqnarray}
   (E-\widehat{H}_0)|\psi\rangle=\lambda \widehat{H}_1|\psi\rangle~.
   \label{EH12}
 \end{eqnarray}
 To obtain $|\psi \rangle$, we have to divide equation (\ref{EH12}) by ($E-\widehat{H}_0$) on both sides. For that we will first examine $(E-\widehat{H}_0)^{-1}$ to see if it is well behaved. Since the operator $(E-\widehat{H}_0)^{-1}$ is a function of the operator $\widehat{H}_0$, it can be written in terms of projectors of $\widehat{H}_0$ as 
 \begin{eqnarray}
   (E-\widehat{H}_0)^{-1}=(E-\widehat{H}_0)^{-1}\sum_{k\alpha}|k\alpha)( k\alpha |~.
 \end{eqnarray}
Here the summation runs over the whole eigen set of $\widehat{H}_0$. That is, $k$ takes $l$ as well as $n$ values. In terms of its eigen values it can be written as
 \begin{eqnarray}
   (E-\widehat{H}_0)^{-1}=\sum_{k\alpha}\frac{1}{(E-E_k)}|k\alpha)( k\alpha |~.
\label{EH01}
 \end{eqnarray}

 However, if any of the unperturbed eigenvalues $E_k$ should equal to the exact eigenvalue $E$, the resultant expression is meaningless. In that situation, one or more denominators vanish, showing that $E-\widehat{H}_0$ lacks an inverse. The terms $k = n$ in Eq. (\ref{EH01}) are substantial (and these terms diverge at $\lambda= 0$, where $E = E_n$), even if none of the denominators vanishes, as the precise eigenvalue $E$ is probably near to the unperturbed eigenvalue $E_n$.

 We define a new operator $\widehat{R}$, which is Eq. (\ref{EH01}) with the elements $k = n$ suppressed, to prevent small or vanishing denominators and to work with operators that are assured to be well defined :
 \begin{eqnarray}
   \widehat{R} =\sum_{l\alpha}\frac{1}{(E-E_l)}|l\alpha)(l\alpha |~.
   \label{Rop}
 \end{eqnarray}
 It is possible that there are other small or vanishing denominators in Eq. (\ref{EH01}) other than $E-E_n$. If, for instance, there are nearby unperturbed energy levels $E_k$, the perturbation can cause the precise energy $E$ to move towards or past some of these levels, leading to the emergence of further tiny denominators in Eq. (\ref{EH01}). If the perturbation is large enough, then this will undoubtedly occur. We shall assume for the time being that this does not occur and that Eq. (\ref{Rop}) is free of small denominators. Mathematically, this corresponds to the situation that $\lambda$ is very close to zero. The operator $\widehat{R}$ satisfies

 \begin{eqnarray}
   \widehat{P}\widehat{R}=\widehat{R}\widehat{P}=0~~~,~~~~~~~~\widehat{Q}\widehat{R}=\widehat{R}\widehat{Q}=\widehat{R}~~~~\textrm{and}~~~~~~\widehat{R}(E-\widehat{H}_0)=(E-\widehat{H}_0)\widehat{R}=\widehat{Q}~.
   \label{PRRP}
 \end{eqnarray}
 Further, we have
 \begin{eqnarray}
  |\psi\rangle=\widehat{P}|\psi\rangle+\widehat{Q}|\psi\rangle~.
  \label{psiPRQ}
\end{eqnarray}
 $\widehat{P}|\psi\rangle$ and $\widehat{Q}|\psi\rangle$ are known as the "easy" and "hard" parts of the exact eigenket $|\psi\rangle$, respectively.  It may be noted that $\widehat{P}|\psi\rangle$ is an eigen function of $\widehat{H}_0$ with eigen value $\epsilon_n$.
 Returning to Eq. (\ref{EH12}), we multiply by $\widehat{R}$ using Eq. (\ref{PRRP}). This gives
\begin{eqnarray}
\widehat{R}(E-\widehat{H}_0)|\psi\rangle=\lambda \widehat{R}\widehat{H}_1|\psi\rangle~~~~~~\Rightarrow~\widehat{Q}|\psi\rangle=\lambda \widehat{R}\widehat{H}_1|\psi\rangle~.
\end{eqnarray}
This expresses the "hard" part of the exact eigenstate $\widehat{Q}|\psi\rangle$ in terms of the exact eigenstate $|\psi\rangle$. To make this more useful, we first add $\widehat{P} |\psi\rangle$ to both sides, resulting in
\begin{eqnarray}
  |\psi\rangle=\widehat{P}|\psi\rangle+\lambda \widehat{R}\widehat{H}_1|\psi\rangle~.
  \label{psiR}
\end{eqnarray}
This represents the exact eigenket $|\psi\rangle$ in terms of its projections onto $\mathcal{H}_ n$ and $\mathcal{H}^{\perp}_n$, the second term being a small correction. As $|\psi\rangle$ occurs on the right hand side, it is not an explicit solution for $|\psi\rangle$, but it can be turned into a power series solution using a successive approximation method.

Notably, if we ignore the second term in Eq. (\ref{psiR}), which is of order of $\lambda$, we get $|\psi\rangle = \widehat{P} |\psi\rangle$, which is correct to the lowest order. To improves the approximation, we substitute the left side of Eq. (\ref{psiR}) into the second term on the right side, yielding
\begin{eqnarray}
  |\psi\rangle=\widehat{P}|\psi\rangle+\lambda \widehat{R}\widehat{H}_1\widehat{P}|\psi\rangle+\lambda^2 \widehat{R}\widehat{H}_1\widehat{R}\widehat{H}_1|\psi\rangle~,
  \label{psiPRH}
\end{eqnarray}
and remains an exact equation. Except for the last term, all terms on the right hand side include the "easy" component $\widehat{P}|\psi\rangle$ of the precise eigenstate, and only the final term, which is second order in the perturbation, involves $|\psi\rangle$. We have an expression for $|\psi\rangle$ in terms of $\widehat{P}|\psi\rangle$ that is valid to first order in $\lambda$ if we neglect that term. Continuing in this manner, the left hand side of Eq. (\ref{psiR}) can be substituted into the last term in Eq. (\ref{psiPRH}), therefore pushing the hard term to fourth order in $\lambda$, and so on. As a consequence, a formal power series for the solution $|\psi\rangle$ in terms of the easy component $\widehat{P}|\psi\rangle$ is obtained.

Substituting Eq. (\ref{PRRP}) in Eq. (\ref{psiPRH}), and neglecting terms with higher powers of $\lambda$, we obtain
\begin{eqnarray}
|\psi\rangle=\widehat{P}|\psi\rangle+\lambda \frac{\widehat{Q}}{(E-\widehat{H}_0)}\widehat{H}_1\widehat{P}|\psi\rangle~.
\label{FPT}
\end{eqnarray}
This is the approximation for the eigenstate of $\widehat{H}$ upto the first power in $\lambda$. This is an important result of the first order perturbation theory.

Let us apply the above result to the neutron-proton system. In this case, we have two degenerate states; $|n)$ and $|p)$ of $\widehat{H}_0$ with energy $E_0=\sqrt{k^2+M_0^2}$. Now, in the presence of the CSB terms in the Hamiltonian,
two non-degenerate states; $|n\rangle$ and $|p\rangle$  result with energies given by Eqs. (\ref{HNP1}) and (\ref{HNP2}), respectively.  Therefore, with the help of Eq. (\ref{FPT}), we have

\begin{eqnarray}
|n\rangle=\widehat{P}|n\rangle+\lambda\frac{\widehat{Q}}{(E-\widehat{H}_0)}\widehat{H}_1\widehat{P}|n\rangle~~~~\mbox{and}~~~~|p\rangle=\widehat{P}|p\rangle+\lambda\frac{\widehat{Q}}{(E-\widehat{H}_0)}\widehat{H}_1\widehat{P}|p\rangle~.
\end{eqnarray}
For $\lambda=0$ in the above equations, we have $|n\rangle=\widehat{P}|n\rangle$ and  $|p\rangle=\widehat{P}|p\rangle$. However, $\lambda=0$ corresponds to the  situation that there are no CSB terms present in the Hamiltonian, thus giving $|n)=\widehat{P}|n\rangle$ and  $|p)=\widehat{P}|p\rangle$. Hence,  $|n\rangle$ and $|p\rangle$ can be written as
\begin{eqnarray}
  |n\rangle=|n)+\lambda\frac{Q}{(E-\widehat{H}_0)}\widehat{H}_1|n)~~~\textrm{and}~~~|p\rangle=|p)+\lambda\frac{Q}{(E-\widehat{H}_0)}\widehat{H}_1|p)~,
  \label{NPket}
\end{eqnarray}
where $\widehat{Q}$ is the projection operator and that
\begin{eqnarray}
\widehat{Q}=I-\widehat{P}~~~~~~~\Rightarrow \widehat{Q}=1-|n)(n|-|p)(p|~.
\end{eqnarray}
Taking the lowest-order perturbation in $\widehat{H}_1$, suppressing indices and setting ${\bf k}=0$, one obtains from Eqs. (\ref{HNP1}) and (\ref{HNP2})

 \begin{eqnarray}
   M_n-M_p=(n|\widehat{H}_1|n)-(p|\widehat{H}_1|p)~,
   \label{MNMP}
 \end{eqnarray}
 which gives the energy difference between the physical states of neutron and proton obtained by the presence of CSB perturbation Hamiltonian. The above equation can be expressed in terms of the 
 commutator $[\widehat{H}, \widehat {P}_{cs}]$ by noting that $\widehat{P}_{cs}^\dagger[\widehat{H},\widehat {P}_{pc}]=\widehat {P}_{cs}^\dagger \widehat{H}\widehat {P}_{cs}-\widehat{H}$ and $|p)=\widehat {P}_{cs}|n)$ as
  \begin{align}
  M_n-M_p&=(n|\widehat{H}_1|n)-(p|\widehat{H}_1|p)~,\nonumber\\
  &=(n|\widehat{H}_1|n)-(n|\widehat {P}_{CS}^\dagger \widehat{H}_1\widehat {P}_{CS}|n)~,\nonumber\\
  &=-(n|\widehat {P}_{CS}^\dagger [\widehat{H}, \widehat {P}_{CS}]|n)~,\nonumber\\
  &=(p|\widehat {P}_{CS}^\dagger [\widehat{H}, \widehat {P}_{CS}]|p)~.
\end{align} 
Assume that the only CSB impact is the difference in quark masses $m_d-m_u$. The wave functions of the bare and physical states are identical, hence $\widehat{H}_1$ commutes with $\widehat{Q}$. In actuality, variations between the bare and physical wave functions are caused by quark kinetic energy and electromagnetic interactions. Now consider the action of $\widehat {P}_{CS}$ on a physical neutron state :
\begin{eqnarray}
  \widehat {P}_{CS}|n\rangle=|p)+\frac{1}{(M_0-\widehat{H}_0)}\widehat {P}_{CS}\widehat{Q} \widehat{H}_1|n)
  =|p)+\frac{1}{(M_0-\widehat{H}_0)}\widehat{Q} \widehat {P}_{CS} \widehat{H}_1|n)~.
\end{eqnarray}
With the help of the second equation of (\ref{NPket}), the above equation can be written as
\begin{eqnarray}
  \widehat {P}_{CS}|n\rangle=|p\rangle+\frac{\widehat{Q}}{(M_0-\widehat{H}_0)}\left( \widehat {P}_{CS}\widehat{H}_1\widehat{P}_{CS}^\dagger - \widehat{H}_1\right)|p)
  \label{pnstate1}
\end{eqnarray}

or

\begin{eqnarray}
  \widehat {P}_{CS}|n\rangle=|p\rangle+\frac{\widehat{Q}}{(M_0-\widehat{H}_0)} [\widehat{P}_{CS}, \widehat{H}_1]|n)
  \label{pnstate2}
\end{eqnarray}

We can similarly express the operation of $\widehat {P}_{CS}$ on the physical proton state. Eq. (\ref{pnstate2}) states that $\widehat {P}_{CS}$ changes a physical neutron state into a proton state only with inclusion of CSB terms.

\setcounter{equation}{0}
\section{Densities in Cylindrical Basis}\label{DENSITIES}
In the case of the axially symmetric even-even nuclei 
with  z-axis as symmetry axis, the third
component of the total angular momentum $j_z$ is a good quantum number for the 
representation of a single-particle state. 
Here, we introduce the axial coordinate representation $(r\phi z)$
for the HFB wave functions. If $\Omega_k$ is the eigenvalue of $j_z$, associated with 
the single-particle state k, then we represent the quasiparticle wave functions for $\Omega_k>0$ as
\begin{eqnarray}\label{eq:QPWF1}
  \begin{pmatrix} \varphi_k(rst) \\ \psi_k(rst) \end{pmatrix}
  = 
  \begin{pmatrix} \varphi_k^+(rzt) \\ \psi_k^+(rzt)  \end{pmatrix}
  e^{i\Lambda^{-}\phi} \chi_{1/2}(s) +
  \begin{pmatrix} \varphi_k^-(rzt) \\ \psi_k^-(rzt) \end{pmatrix}
    e^{i\Lambda^{+}\phi} \chi_{-1/2}(s)~,
\end{eqnarray}
where
\begin{equation*}
 \Lambda^{\pm 
}=\Big(\Omega_k\pm\frac{1}{2}\Big) .
\end{equation*} 

In the explicit derivation of densities and currents in the cylindrical coordinates, the following 
expressions are quite useful,
\begin{enumerate}
 \item We have for the derivatives of spin matrices
\begin{align} \nonumber
(\partial_{\phi}\widehat{\boldsymbol\sigma}_{s's})&=\partial_{\phi}\Big(\widehat{\sigma}^{r}_{s's}\widehat{\boldsymbol e}_{r
}+\widehat
{ \sigma}^{\phi}_{s's}\widehat{\boldsymbol e}_{\phi}+\widehat{\sigma}^{z}_{s's}\widehat{\boldsymbol e}_{z}  \Big)~,\\\nonumber 
&=\Big(\widehat{\sigma}^{r}_{s's}\widehat{\boldsymbol e}_{\phi}+(\partial_{\phi}\widehat{\sigma}^{r}_{s's})\widehat{\boldsymbol e}_{r}
-\widehat{\boldsymbol e}
_{r}\widehat{\sigma}^{\phi}_{s's}  
+(\partial_{\phi}\widehat{\sigma}^{\phi}_{s's})\widehat{\boldsymbol e}_{\phi}+0\Big)+\widehat{\boldsymbol\sigma}^{}_{s's}\partial_{
\phi}~,
\\\nonumber
&=\Bigg\{\begin{pmatrix} 0 & e^{-i\phi} \\ e^{i\phi} & 
0\end{pmatrix}\widehat{\boldsymbol e}_{\phi}+(\partial_{\phi}\begin{pmatrix} 0 & e^{-i\phi} \\ e^{i\phi} & 
0\end{pmatrix})\widehat{\boldsymbol e}_ {r}~.\\\nonumber 
&-\widehat{\boldsymbol e}
_{r}\begin{pmatrix} 0 &-i e^{-i\phi} \\i e^{i\phi} & 
0\end{pmatrix} +(\partial_{\phi}\begin{pmatrix} 0 &-i e^{-i\phi} \\i e^{i\phi} & 
0\end{pmatrix} )\widehat{\boldsymbol e}_{\phi}\Bigg\}+\widehat{\boldsymbol\sigma}^{}_{s's}\partial_{\phi}~,
\\\nonumber 
&=\Bigg\{\begin{pmatrix} 0 & e^{-i\phi} \\ e^{i\phi}&0\end{pmatrix}\widehat{\boldsymbol e}_{\phi}  +\begin{pmatrix} 
0 
& 
-ie^{-i\phi} \\ ie^{i\phi} &0\end{pmatrix}\widehat{\boldsymbol e}_ {r} \\\nonumber & -\begin{pmatrix} 0 &-i e^{-i\phi} 
\\i e^{i\phi} &0\end{pmatrix}\widehat{\boldsymbol e}_{r}  
+\begin{pmatrix} 0 &- e^{-i\phi} \\- e^{i\phi} & 
0\end{pmatrix}\widehat{\boldsymbol e}_{\phi}\Bigg\}+\widehat{\boldsymbol\sigma}^{}_{s's}\partial_{\phi}~,
\\
&= \widehat{\boldsymbol\sigma}^{}_{s's}\partial_{\phi}~.
\end{align}
where $\widehat{{\bm\sigma}}^r_{ss'},\widehat{{\bm\sigma}}^\phi_{ss'},\widehat{{\bm\sigma}}^z_{ss'}$ are Pauli spin matrices in cylindrical polar coordinate system and $\widehat{\sigma}^0_{ss'}=\delta _{ss'}$.
\item Further, the components of the   operator   
$(\boldsymbol{\nabla}\times\widehat{\boldsymbol\sigma})= 
({\boldsymbol\nabla}\times{\widehat{\boldsymbol\sigma}})_{r}\widehat{\boldsymbol e}_{r}+(\widehat{\boldsymbol\nabla}\times{\widehat{\boldsymbol\sigma}})_{\phi}
\widehat{\boldsymbol e
}_{\phi} +({\boldsymbol\nabla} \times{\widehat{\boldsymbol\sigma}})_{z} \widehat{\boldsymbol e}_{z}$, are given by
\begin{align}   
(-i)({\boldsymbol\nabla}\times{\widehat{\boldsymbol\sigma}})_{r}&=(-i)\widehat{\sigma}_{z}\frac{\partial_{\phi}}{r}+\frac{1
}{2}(\widehat{\sigma}_ {+}e^{-i\phi}-\widehat{\sigma}_{-}e^{i\phi})\partial_{z}~,\\
({\boldsymbol\nabla}\times\widehat{{\boldsymbol\sigma}})_{\phi}&=\frac{1}{2}(\widehat{\sigma}_{+}e^{-i\phi}+\widehat{\sigma}
_{-} e^{
i\phi })\partial_ {z}-\widehat{\sigma}_{z}\partial_{r}~,\\
(-i)({\boldsymbol\nabla}\times{\widehat{\boldsymbol\sigma}})_{z}&=-\frac{1}{2}\widehat{\sigma}_{+}e^{-i\phi}\Big(\partial_{
r}
-i\frac{\partial_{\phi}}{r} 
\Big)+\frac{1}{2}\widehat{\sigma}_{-}e^{i\phi}\Big(\partial_{r}+i\frac{\partial_{\phi}}{r} 
\Big) ~,
\end{align}
where 
\begin{eqnarray*} 
\widehat{\sigma}_{+}=\begin{pmatrix} 0 & 2\\0&0\end{pmatrix};
\qquad \textrm{and} \qquad \widehat{\sigma}_{-}=\begin{pmatrix} 0 & 0\\2&0\end{pmatrix}~.
\end{eqnarray*}
\item For the Laplacian, we have
\begin{align}\nonumber
  \nabla^{2}&= \frac{1}{r}\frac{\partial}{\partial {r}}\Big(r\frac{\partial}{\partial {r}} \Big) 
  +\frac{1}{r^2}\frac{\partial^{2}}{\partial\phi^{2}}+ \frac{\partial^{2}}{\partial{z}^{2}} ~,
  \\
  \nabla^{2}(\widehat{\boldsymbol\sigma} f)&= \widehat{\boldsymbol\sigma}\Bigg( \frac{1}{r}\frac{\partial}{\partial 
{r}}\Big(r\frac{\partial}{\partial {r}} \Big) +\frac{\partial^{2}}{\partial{z}^{2}}  \Bigg) 
f+\frac{1}{r^2}\frac{\partial^{2}}{\partial\phi^{2}}(\widehat{\boldsymbol\sigma} f) ~,
\end{align}
and it can be shown that

\begin{eqnarray}
 \frac{1}{r^2}\frac{\partial^{2}}{\partial\phi^{2}}(\widehat{\boldsymbol\sigma} f) 
=\frac{1}{r^2}\frac{\partial^{2}}{\partial\phi^{2}}\Big[\widehat{\boldsymbol e}_{r}\widehat{\sigma}_{r}f+\widehat{\boldsymbol 
e}_{\phi}
\widehat{\sigma}_{\phi}f+\widehat{\boldsymbol e}_{z}
\widehat{\sigma}_{z}f\Big]=\widehat{\boldsymbol\sigma}\Big[\frac{1}{r^2}\frac{\partial^{2}f}{\partial\phi^{2}}\Big]~.
\end{eqnarray}
Therefore 
\begin{eqnarray*}
  \nabla^{2}(\widehat{\boldsymbol\sigma} f)= \widehat{\boldsymbol\sigma}\nabla^{2} f~.
\end{eqnarray*}
\end{enumerate}

The density matrices in the spin and isospin spaces are expressed as the linear combination of spin and isospin Pauli matrices, where the notation used in the Ref.\cite{Per04} has been followed.

\begin{enumerate}
\item{\bf Particle Density}

The particle-hole (ph) isoscalar and isovector non-local densities are expressed in terms of the quasiparticle 
wave functions and isospin Pauli matrices as
  \begin{align*}   
\rho_m(\boldsymbol{r},\boldsymbol{r}')&=\sum_{stt'}\widehat{\rho}(\boldsymbol{r}st,\boldsymbol{r}'st')\widehat{\tau}^{m}_{t't}~,\\
\rho_m(\boldsymbol{r},\boldsymbol{r}')&=\sum_{stt'~k}(4tt')\psi^{*}_{k}(\boldsymbol{r}-s-t)\psi^{}_{k}(\boldsymbol{{r}'}-s-t')\widehat{
\tau}^{m}_{
t't}\\\nonumber
&~~~~~+\sum_{stt'~\bar{k}}(4tt')\psi^{*}_{\bar{k}}(\boldsymbol{r}-s-t)\psi^{}_{\bar{k}}(\boldsymbol{{r}'}-s-t')\widehat{\tau
}^{m}_ {
t't}~.
\end{align*}
Using the cylindrical expansion of the quasiparticle  wave function Eq. (\ref{eq:QPWF1}), we obtain
 \begin{align} \nonumber  
\rho_m(\boldsymbol{r},\boldsymbol{r}')&=\sum_{tt'~k>0}(4tt')\Bigg[\psi^{+*}_k(rz-t) 
\psi^+_k(r'z-t') 
[e^{-i\Lambda^-(\phi-\phi')}+e^{i\Lambda^-(\phi-\phi')}] \\
&~~~~~~~~~~~~~~~~~~+\psi_k^{-*}(rz-t) 
\psi_k^-(r'z-t')[e^{-i\Lambda^+(\phi-\phi')}+e^{i\Lambda^+(\phi-\phi')}]\Bigg]\widehat{\tau}^{m 
}_{t't}~,
\end{align}
where ${\tau}^0_{tt'}=\delta _{tt'}$ and ${\tau}^1_{tt'},{\tau}^2_{tt'},{\tau}^3_{tt'}$ are the isospin Pauli matrices.

\item{\bf Vector Spin  Density}

The ph spin isoscalar and isovector densities are expressed as
\begin{align}\nonumber     
\boldsymbol{s}_m(\boldsymbol{r},\boldsymbol{r}')&=\sum_{ss'tt'}\widehat{\rho}(\boldsymbol{r}st,\boldsymbol{r}s't')\widehat{\boldsymbol\sigma}_{
s's }\widehat{\tau}^{m}_{ t't } ~,\nonumber \\\nonumber 
&=\sum_{ss'tt'~k}(16ss'tt')\psi^{*}_{k}(\boldsymbol{r}-s-t)\psi^{}_{k}(\boldsymbol{{r}'}-s'-t')\widehat{
\boldsymbol\sigma } _{
s's}\widehat{\tau}^{m}_{t't}~,\nonumber \\\nonumber 
&=\sum_{ss'tt'~k}(16ss'tt')\psi^{*}_{k}(\boldsymbol{r}-s-t)\psi^{}_{k}(\boldsymbol{{r}'}-s'-t')\nonumber \\
&~~~~~~~~~~~~~~~~~~~~~~~~~~~~~~~~~~~~~~~~~~~~~~~~~~~~~\Big(\widehat{e
}_{ r } {\sigma }^{r} _ {s's}+\widehat{e}_{\phi}{\sigma }^{\phi} _ {s's}
+\widehat{e}_{z} {\sigma }^{z} _ {s's}\Big)\widehat{\tau}^{m}_{t't}~,\nonumber \\
&= \sum_{tt'}(4tt') 
[ s_r({\boldsymbol r}t,{\boldsymbol r}'t')\widehat{\boldsymbol{e}}_r + s_\phi({\boldsymbol r}t,{\boldsymbol r}'t')\widehat{\boldsymbol{e}}_\phi+ 
s_z({\boldsymbol r}t,\boldsymbol{r'}t')\widehat{\boldsymbol{e}}_z]
\widehat{\tau}^m_{t't} ~,\\
&\hspace{-1.5cm}\textrm{where}\nonumber\\
s_r({\bf r}t,{\boldsymbol r}'t')&=\sum_{ss'~k}(4ss')\psi^{*}_{k}(\boldsymbol{r}-s-t)\psi^{}_{k}(\boldsymbol{{r}'}
-s'-t'){\sigma }^{r} _ {s's}~,\nonumber\\ 
&=\sum_{ss'~k>0}(4ss')\psi^{*}_{k}(\boldsymbol{r}-s-t)\psi^{}_{k}(\boldsymbol{{r}'}
-s'-t'){\sigma }^{r} _ 
{s's}\nonumber\\
&~~~~~~~~~~~~+\sum_{ss',\bar{k}}(4ss')\psi^{*}_{\bar{k}}(\boldsymbol{r}-s-t)\psi^{}_{\bar{k}}(\boldsymbol{{r}'
}-s'-t'){\sigma }^{r} _ {s's}~,\nonumber \\
&=\sum_{k>0}\Big(\psi^{+*}_k(rz-t)\psi_k^-(r'z-t') 
[e^{i\Lambda^+(\phi-\phi')}-e^{-i\Lambda^+(\phi-\phi')}]\nonumber\\
&~~~~~~~~~~~~+\psi^{-*}_k(rz-t)\psi_k^+(r'z-t') 
[e^{i\Lambda^-(\phi-\phi')}- e^{-i\Lambda^-(\phi-\phi')}]
\Big)~,\\
 s_\phi({\boldsymbol r}t,{\bf  
r}'t')&=\sum_{ss'~k}(4ss')\psi^{*}_{k}(\boldsymbol{r}-s-t)\psi^{}_{k}(\boldsymbol{r }'
-s'-t'){\sigma }^{\phi} _ {s's}~,\nonumber\\
&= (i)\sum_{k>0}\Big\{\psi^{+*}_k(rz-t)\psi_k^-(r'z-t') 
[e^{-i\Lambda^+(\phi-\phi')}+e^{i\Lambda^+(\phi-\phi')}]\nonumber\\
&~~~~~~~~~~~~-\psi^{-*}_k(rz-t)\psi_k^+(r'z-t') 
[e^{-i\Lambda^-(\phi-\phi')}+e^{i\Lambda^-(\phi-\phi')}]
\Big\}~,\\
s_z({\boldsymbol r}t,\boldsymbol{r'}t')&=\sum_{ss'~k}(4ss')\psi^{*}_{k}(\boldsymbol{r}-s-t)\psi^{}_{k}(\boldsymbol{ 
{r}'} -s'-t'){\sigma }^{z} _ {s's}~,\nonumber \\
&=\sum_{k>0}\Big\{\psi^{+*}_k(rz-t)\psi_k^+(r'z-t') 
[e^{i\Lambda^-(\phi-\phi')}-e^{-i\Lambda^-(\phi-\phi')}]\nonumber\\
&~~~~~~~~~~~~-\psi^{-*}_k(rz-t)\psi_k^-(r'z-t') 
[e^{i\Lambda^+(\phi-\phi')}- e^{-i\Lambda^+(\phi-\phi')}]
\Big\} ~.
\end{align}

\end{enumerate}
\end{appendices}
\bibliographystyle{unsrt}
\bibliography{isb}
\end{document}